\documentclass{article}

\usepackage{amsmath,amssymb}
\usepackage[margin=1in]{geometry}
\usepackage{graphicx}
\usepackage{float}

\title{Reduced order prediction of rare events in unidirectional nonlinear water waves}
\author{Will Cousins (willcousins@gmail.com)  and  Themistoklis P. Sapsis (sapsis@mit.edu) \\ Department of Mechanical Engineering \\ Massachusetts Institute of Technology \\ 77 Massachusetts Ave., Cambridge, MA, USA}
\date{}

\begin{document}


\maketitle

\begin{abstract}
We consider the problem of short-term prediction of rare, extreme water waves in unidirectional fields, a critical topic for ocean structures and naval operations. One possible mechanism for the occurrence of such rare, unusually-intense waves is nonlinear wave focusing. Recent results have demonstrated that random localizations of energy, induced by the dispersive mixing of different harmonics, can grow significantly due to localized nonlinear focusing. Here we show how the interplay between i) statistical properties captured through linear information such as the waves power spectrum and ii) nonlinear dynamical properties of focusing localized wave groups defines a critical length scale associated with the formation of extreme events. The energy that is locally concentrated over this length scale acts as the `trigger' of  nonlinear focusing for wave groups and the formation of subsequent rare events. We use this property to develop inexpensive, short-term predictors of large water waves. Specifically, we show that by merely tracking the energy of the wave field over the critical length scale allows for the robust, inexpensive prediction of the location of intense waves with a prediction window of 25 wave periods. We demonstrate our results in numerical experiments of unidirectional water wave fields described by the Modified Nonlinear Schrodinger equation. The presented approach introduces a new paradigm for understanding and predicting intermittent and localized events in dynamical systems characterized by uncertainty and potentially strong nonlinear mechanisms.


\end{abstract}

\section{Introduction}
\label{sec:introduction}

Understanding and predicting nonlinear water waves, especially those characterized by large magnitude, is one of the most challenging topics for ocean engineering both because of the catastrophic impact they can have on ocean engineering structures (e.g. for ships and offshore platforms) and naval operations, but also because of the serious lack of specialized mathematical tools for the analysis of the underlying physics \cite{dysthe2008,akhmediev2010,muller2005,xiao2013}. This is because nonlinear water wave dynamics are characterized both by the existence of inherent uncertainty (expressed in the form of phase uncertainty between different Fourier modes) and also in the form of strong nonlinearities and associated energy transfers between modes. The latter can be activated locally and intermittently, leading to unusually high magnitude waves that emerge out of the complex background wave field.

An extreme form of such dynamical evolution is the case of freak or rogue waves, which can be as large as four times the standard deviation of the surrounding wave field \cite{onorato2005_PoF,dysthe2008}. However, waves of even lower magnitude (but larger compared with the surrounding wave field) can still occur with large probability, causing the heavy-tailed statistics for water wave elevations \cite{onorato2005_PoF}. Waves of this magnitude have caused considerable damage to ships, oil rigs, and human life \cite{haver2004,liu2007}. In addition, many naval operations e.g. transfer of cargo between ships moored together in a sea base, landing on aircraft carriers, or path planning of high-speed surface vehicles require short-term prediction of the surrounding wave field.  To make such predictions, unusually high wave elevations must be forecasted reliably and inexpensively.

One mechanism for the occurrence of such rare, unusually-intense wave elevations is nonlinear wave focusing \cite{janssen2003,onorato2002,cousinsSapsis2015_PRL}. For deep water waves a manifestation of this focusing is the well-known Benjamin-Feir instability of a plane wave to small sideband perturbations. This instability, which has also been demonstrated experimentally \cite{chabchoub2011}, generates huge coherent structures by soaking up energy from the nearby field \cite{benjaminFeir1967,zakharov1968,osborne2000}. In \cite{cousinsSapsis2015_PRL}, it was recently demonstrated that even imperfect background conditions, i.e. completely different from the idealized plane wave setup of the Benjamin-Feir instability, can still lead to important wave focusing and rare events. In particular, it was analytically shown and numerically demonstrated for unidirectional wavetrains that there is a critical combination of wave group length scales and amplitudes which will lead to wave focusing and thus unusually high elevations. In contrast to the standard BF mechanism these instabilities have an essentially localized character. 

Random relative phases between different harmonics can potentially lead to the `triggering' of this localized focusing mechanism. Such phase randomness is mainly introduced by the mixing of different harmonics due to their dispersive propagation. Here we show how the interplay between i) statistical properties captured through linear information such as the waves power spectrum and ii) nonlinear focusing properties of localized wave groups and in particular nonlinear focusing properties, defines a critical length scale that is associated with the occurrence of strongly nonlinear interactions and the formation of extreme events. The energy that is locally concentrated over this length scale acts essentially as the `trigger' of nonlinear focusing of wave groups. 

We use this property to develop short-term predictive schemes for the occurrence of large water waves. Specifically, we first demonstrate that using a scale-selection algorithm \cite{lindeberg1998} allows for the partition of the current wave field into a set of wave groups having different length scale and amplitude. A direct computation of the analytical nonlinear stability criterion for focusing of localized wavegroups \cite{cousinsSapsis2015_PRL} allows for a very inexpensive forecast of nonlinear focusing and subsequent growth for each wave group of the field. In a second stage, we demonstrate that merely tracking the energy of the wave field over the critical length scale defined by the interplay between statistics and nonlinearity allows for an even cheaper and robust forecast of upcoming intense nonlinear wave elevations, with a prediction window on the order of 25 wave periods. We demonstrate our results in numerical experiments involving unidirectional water wave fields described by the Modified Nonlinear Schrodinger equation. 

The proposed predictive method reveals and directly utilizes the low-dimensional character for the domain of attraction to these rare water waves. In particular, despite the distribution of the background energy over a wide range of scales, the `trigger' of nonlinear focusing is essentially low-dimensional and to this end it can be used as an inexpensive way to estimate the probability for a rare event in the near future. In addition, the association of the predictor with the energy over a specific length scale gives strong robustness properties against measurement errors and energy in other spatial scales of the wave field. The presented approach introduces a new paradigm for handling spatiotemporal rare events in dynamical systems with inherent uncertainty by providing an efficient description of the `trigger' that leads to those rare events through the careful study of the synergistic action between uncertainty and nonlinearity.

\section{Extreme Events in Envelope Equations}
\label{sec:equations}

In this work, we consider waves traveling on the surface of a fluid of infinite depth. A typical approach for modeling this phenomena is to assume incompressible, irrotational, inviscid flow, which gives Laplace's equation for the velocity potential.  This equation is paired with two boundary conditions: a pressure condition and a kinematic one (a particle initially on the free surface remains so).  This model agrees well with laboratory experiments \cite{wu1998}, and faithfully reproduces the classical $k^{-5/2}$ spectral tail observed in deep water \cite{onorato2002_PRL}.  Although some care is required to numerically deal with the free surface, this fully nonlinear model may be solved numerically with reasonable computational effort, particularly in one space dimension \cite{dommermuth1987,craig1993,dyachenko1996,choi1999}.  However, the presence of the free surface makes analysis of the underlying dynamics challenging.

Thus, in this work we consider approximate equations governing the evolution of the wave envelope, the Nonlinear Schrodinger Equation (NLS) \cite{zakharov1968} and the Modified Nonlinear Schrodinger Equation (MNLS) of \cite{dysthe1979}.  Both NLS and MNLS can be derived via a perturbation approach from the fully nonlinear model under assumptions of small steepness and slow variation of the wave envelope.  Although forms of these equations exist in a full two-dimensional setting, here we consider wave fields varying only in the direction of propagation.  The NLS equation, in nondimensionalized coordinates, reads
\begin{align}
	\frac{\partial u}{\partial t} + \frac{1}{2} \frac{\partial u}{\partial x} + \frac{i}{8} \frac{\partial ^ 2u}{\partial x^2} + \frac{i}{2} |u|^2 u = 0
	\label{eq:NLS}
\end{align}
where $u(x,t)$ is the wave envelope. To leading order the surface elevation is given by $\eta(x,t) = \Re[u(x,t) e^{i(x-t)}]$.  Equation (\ref{eq:NLS}) has been nondimensionalized with $x = k_0 \tilde x, t = \omega_0 \tilde t, u = k_0 \tilde u$, where $\tilde x, \tilde t$, and $\tilde u$ are physical space, time and envelope. $k_0$ is the dominant spatial frequency of the surface elevation, and $\omega_0 = \sqrt{g k_0}$.

Our primary interest in this work is the Modified NLS equation, which is a higher order approximation of the fully nonlinear model,

\begin{align}
  \frac{\partial u}{\partial t} + \frac{1}{2} \frac{\partial u}{\partial x} + \frac{i}{8} \frac{\partial^2 u}{\partial x^2} - \frac{1}{16} \frac{\partial^3 u}{\partial x^3} + \frac{i}{2} |u|^2 u + \frac{3}{2} |u|^2 \frac{\partial u}{\partial x} + \frac{1}{4} u^2 \frac{\partial u^*}{\partial x} + i u \frac{\partial \phi}{\partial x} \Big|_{z=0} = 0 \label{eq:MNLS}
\end{align}
where $\phi$ is the velocity potential and $\partial \phi / \partial x \Big|_{z=0} = -\mathcal{F}^{-1}\left[|k|\mathcal{F}[|u|^2]\right]/2$. $\mathcal{F}$ denotes the Fourier transform.  The MNLS equation has been shown to reproduce laboratory experiments reasonably well \cite{lo1985,goullet2011}.  There are even higher order envelope equations, such as the Broadband Modified NLS Equation (BMNLS) \cite{trulsen1996}.  However, we do not discuss these equations here. Although there are considerable differences between NLS and MNLS, we found minimal differences between simulations of MNLS and BMNLS.  Dysthe et. al. also found similar agreement between MNLS and BMNLS \cite{dysthe2003}.

These envelope equations, as well as the fully nonlinear water wave model, admit periodic plane wave solutions.  Interestingly, these plane wave solutions are unstable to sideband perturbations.  This instability, termed the Benjamin-Feir instability after its discoverers (\cite{benjaminFeir1967}, see also Zakharov \cite{zakharov1968}), has a striking manifestation.  Energy is ``sucked up'' from the nearby field to produce a large amplitude coherent structure, containing a wave 2.4-3 times larger than the surrounding background field \cite{osborne2000}. This behavior is been shown numerically in envelope equations \cite{yuen1978,dysthe1999} as well as the fully nonlinear formulation \cite{henderson1999}.  Furthermore, a number of experiments confirm these numerical predictions \cite{chabchoub2011,chabchoub2012}.

However, in realistic physical settings, the water surface is not merely a plane wave--energy is distributed over a range of frequencies. Thus, in this work, we consider extreme waves emerging out of a background with Gaussian spectra and random phases, that is

\begin{align*}
	u(x,0) = \sum_{-N/2+1}^{N/2} \sqrt{2 \Delta_k F(k\Delta_k)} e^{i(\omega_k x + \xi_k)}, \hspace{0.5in} F(k) = \frac{\epsilon^2}{\sigma \sqrt{2\pi}} e^{\frac{-k^2}{2\sigma^2}}
\end{align*}
where $\xi_k$ are independent, uniformly distributed random phases. Here we adopt the definition of an extreme wave as any instance where $|u| > H_E = 4 \epsilon$ (as defined above $\epsilon$ is the standard deviation of the surface elevation). In these irregular wave fields, it is well known that the critical quantity for extreme event formation is the Benjamin-Feir Index, which is the ratio of the energy level of the field to its bandwidth \cite{janssen2003}.  If the Benjamin-Feir Index is large enough, then nonlinear interactions dominate, leading to the appearance of large amplitude coherent structures and heavy-tailed statistics for the elevation \cite{alber1978,crawford1980,dysthe2003,janssen2003,onorato2005_PoF}. 

We solve the MNLS equation numerically using a Fourier method in space.  The use of periodic boundary conditions is of course artificial, but is a standard convention.  We take our spatial domain to be 128 wavelengths (256$\pi$), large enough to avoid any box-size effects.  We use a 4th order Runge-Kutta exponential time differencing scheme \cite{cox2002,grooms2014}.  This scheme requires evaluation of the function $\phi(z) = (e^z-1)/z$.  Naive computation of $\phi$ can suffer from numerical cancellation error for small z \cite{kassam2005}.  We use a Pade approximation code from the EXPINT software package, which does not suffer from such errors \cite{berland2007}. We use $2^{10}$ Fourier modes with a time step of 0.025; results in this work were insensitive to further refinement in grid size.  

\begin{figure}[H]
	\centering
	\includegraphics[width=0.35\textwidth]{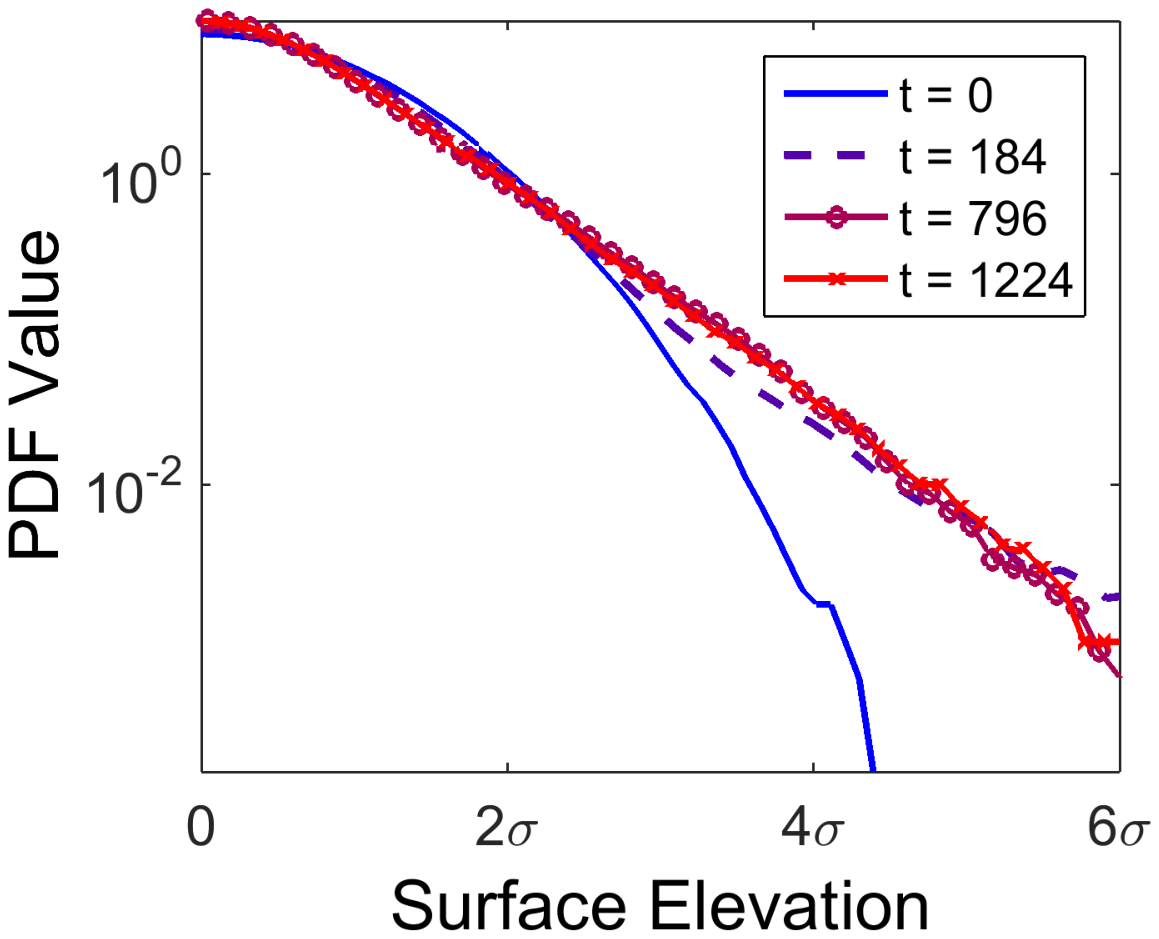} \raisebox{0.15\height}{\includegraphics[width=0.6\textwidth]{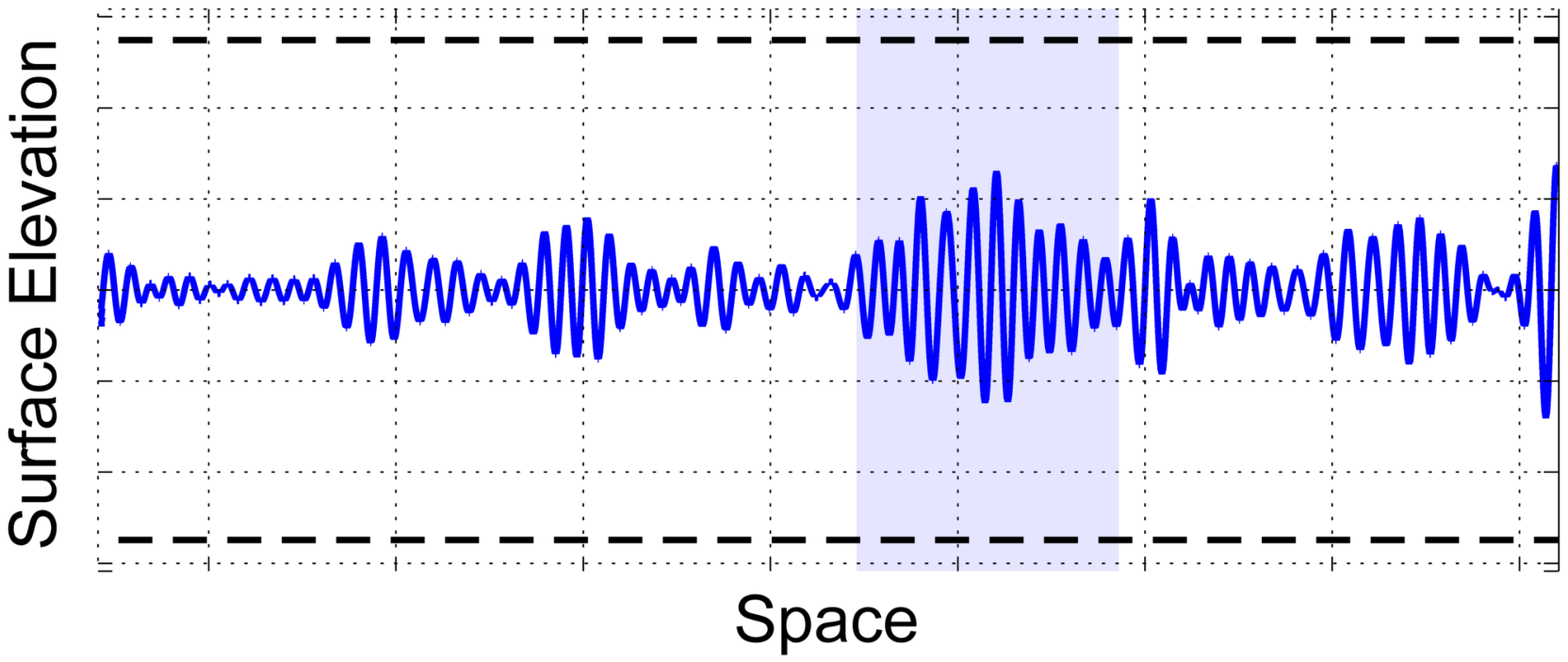}} \\ [1em]
	\includegraphics[width=0.35\textwidth]{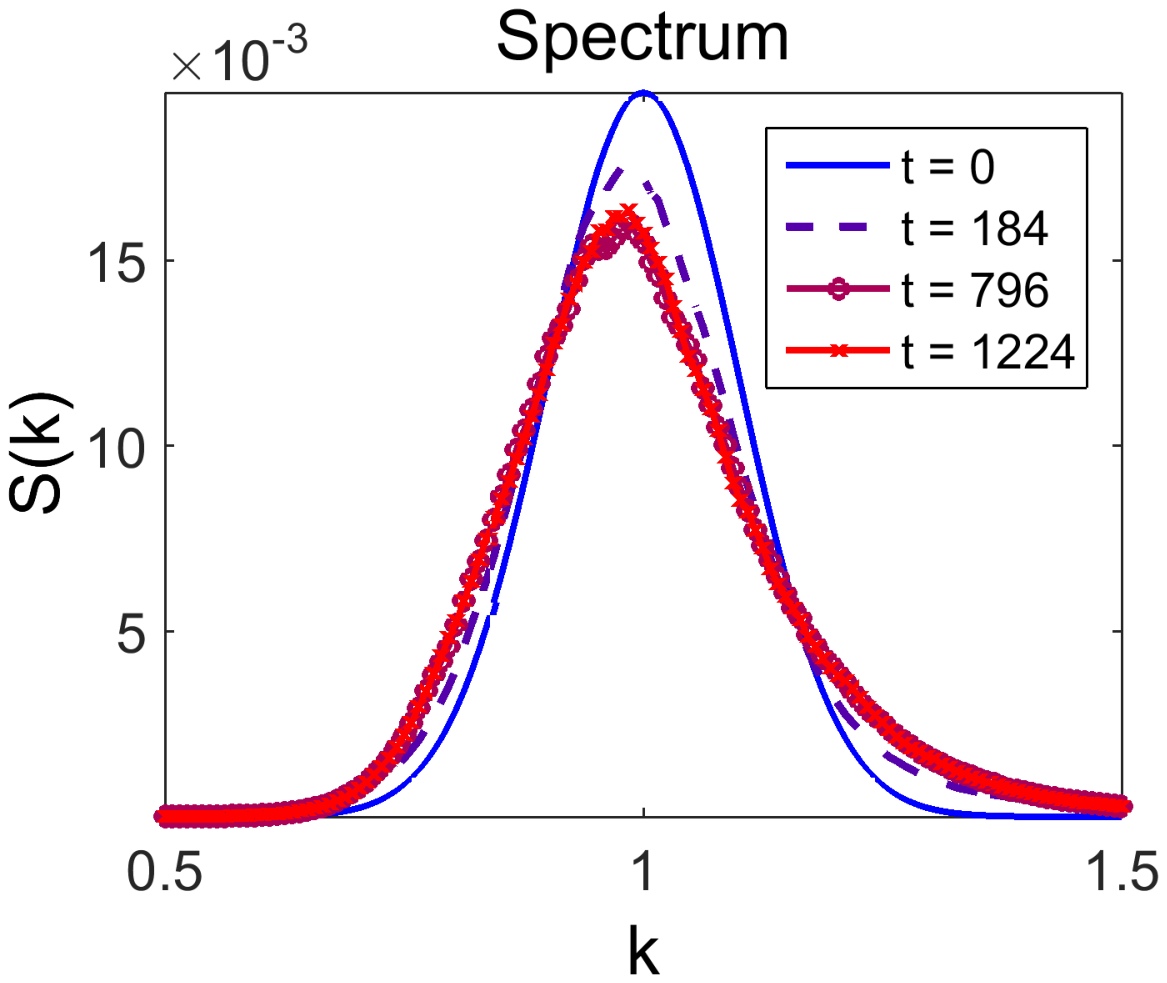} \raisebox{0.15\height}{\includegraphics[width=0.6\textwidth]{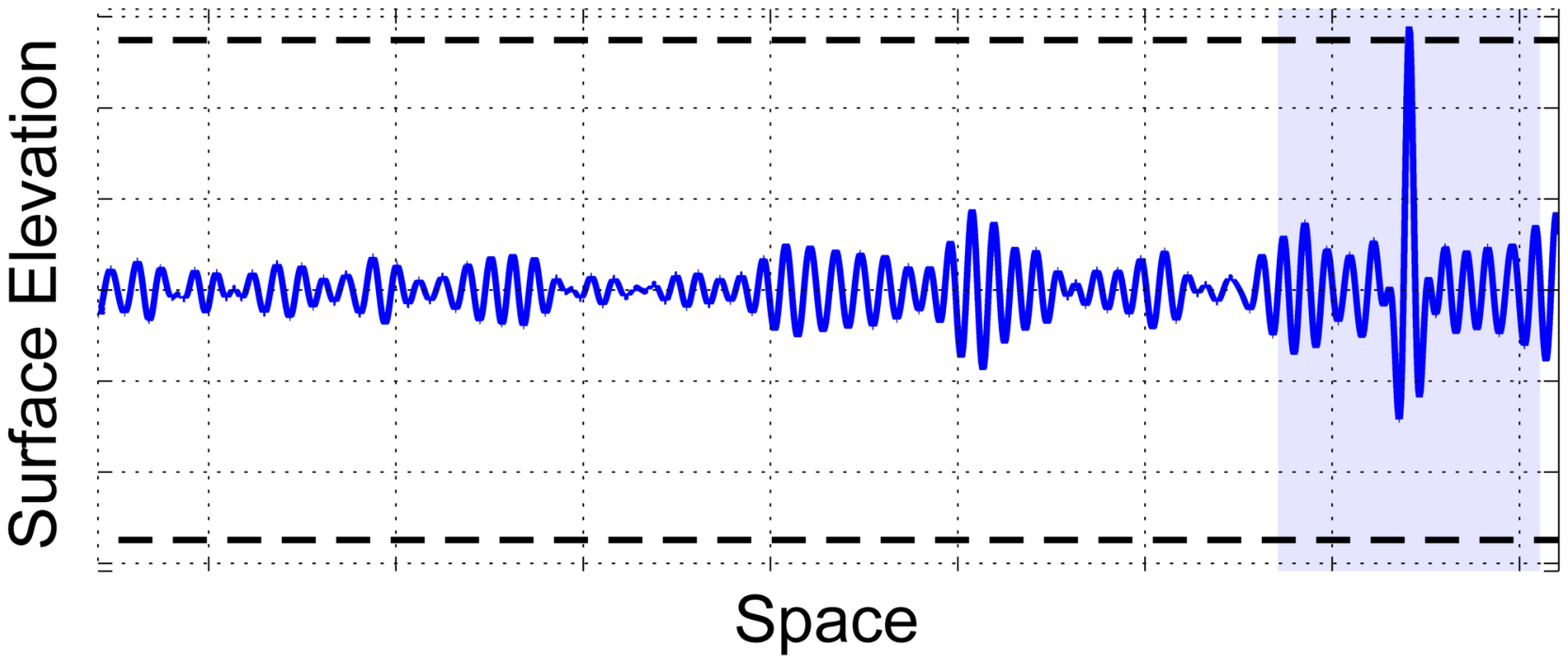}}
	\caption{Top Left: Probability density function of the surface elevation in numerical simulations of MNLS for various times.  Initially the density is Gaussian but eventually develops heavy tails due to nonlinear interactions, which also cause the spectrum to change shape (bottom left).  These heavy tails are due to large amplitude coherent structures (bottom right), which emerge via focusing of localized wave groups (top right).}
	\label{fig:EEIntro}
\end{figure}

\section{Localized Wave Group Evolution}
\label{sec:local}

A large Benjamin-Feir Index indicates that extreme events are more likely than Gaussian statistics would suggest.  However, a large Benjamin-Feir Index does \emph{not} provide any specific information on precisely where an extreme event will occur. Thus, in order to develop a scheme providing precise spatiotemporal predictions, we must develop a more precise indicator than the BFI.  We observed in simulations of high BFI fields, extreme events appear to be triggered by the focusing of localized wave groups (Ruban has also made a similar observation \cite{ruban2013}).  Figure~\ref{fig:EEIntro} displays such an example of extreme event formation by focusing of localized groups.  In this example we see that a localized group focuses, narrowing in width and doubling in amplitude, yielding an extreme event.  

To better understand this mechanism, in \cite{cousinsSapsis2015_PRL} we studied the evolution of isolated wave groups.  We briefly review these results describing the evolution of hyperbolic secant initial data:
\begin{align*}
	u(x,0) = A \text{sech}(x/L)
\end{align*}
Due to the invariance of the envelope equations we take $A$ real with no loss of generality.  We investigated the evolution of such groups as a function of amplitude $A$ and length scale $L$.    In particular, we are interested in whether or not a group will focus and, if it does, the degree by which the group amplitude is magnified.  	Similar questions asked by Adcock et al. in the context of the one-dimensional NLS \cite{adcock2009} as well as NLS and the fully nonlinear model in two dimensions \cite{adcock2012}.

\begin{figure}[H]
	\centering
	\includegraphics[width=0.48\textwidth]{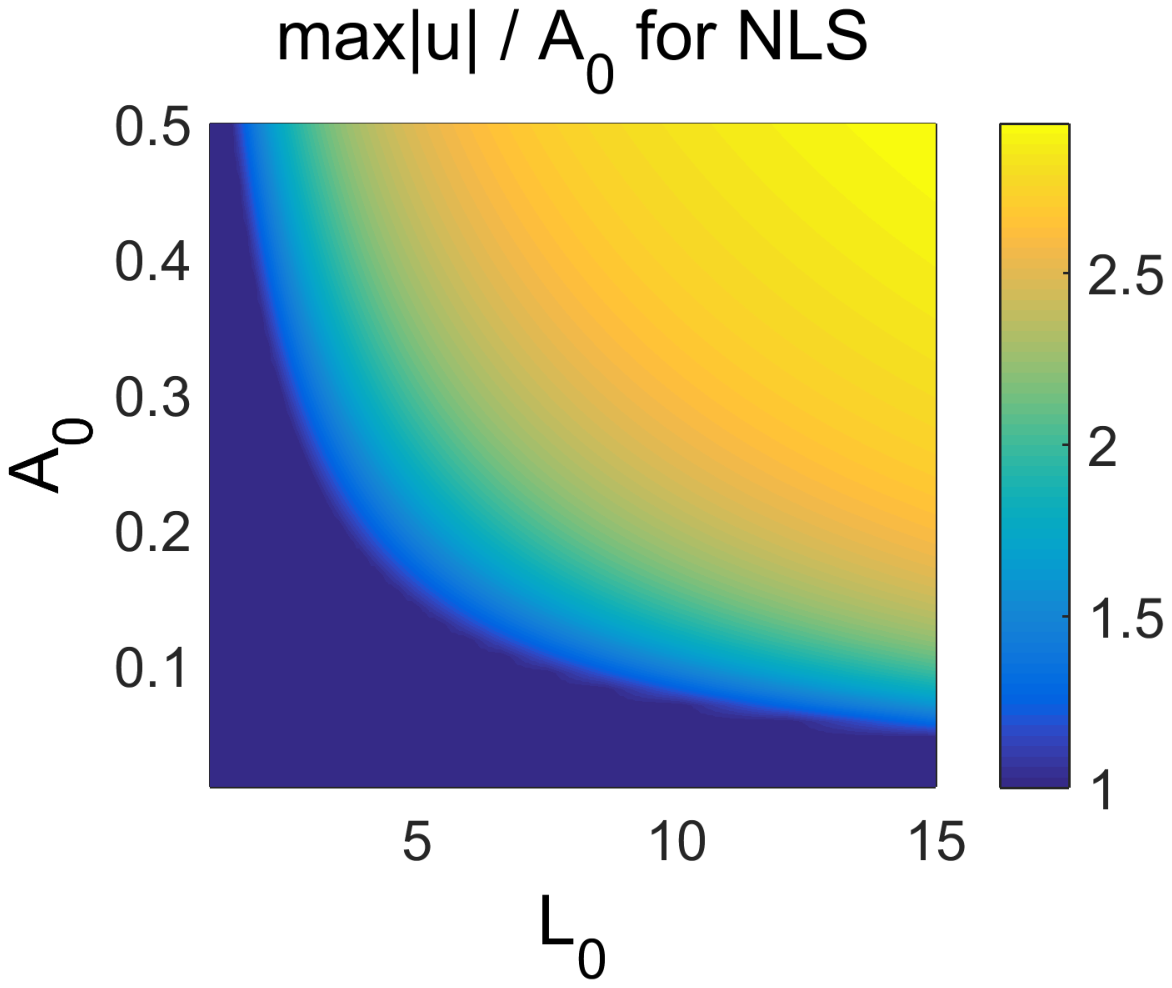} \includegraphics[width=0.48\textwidth]{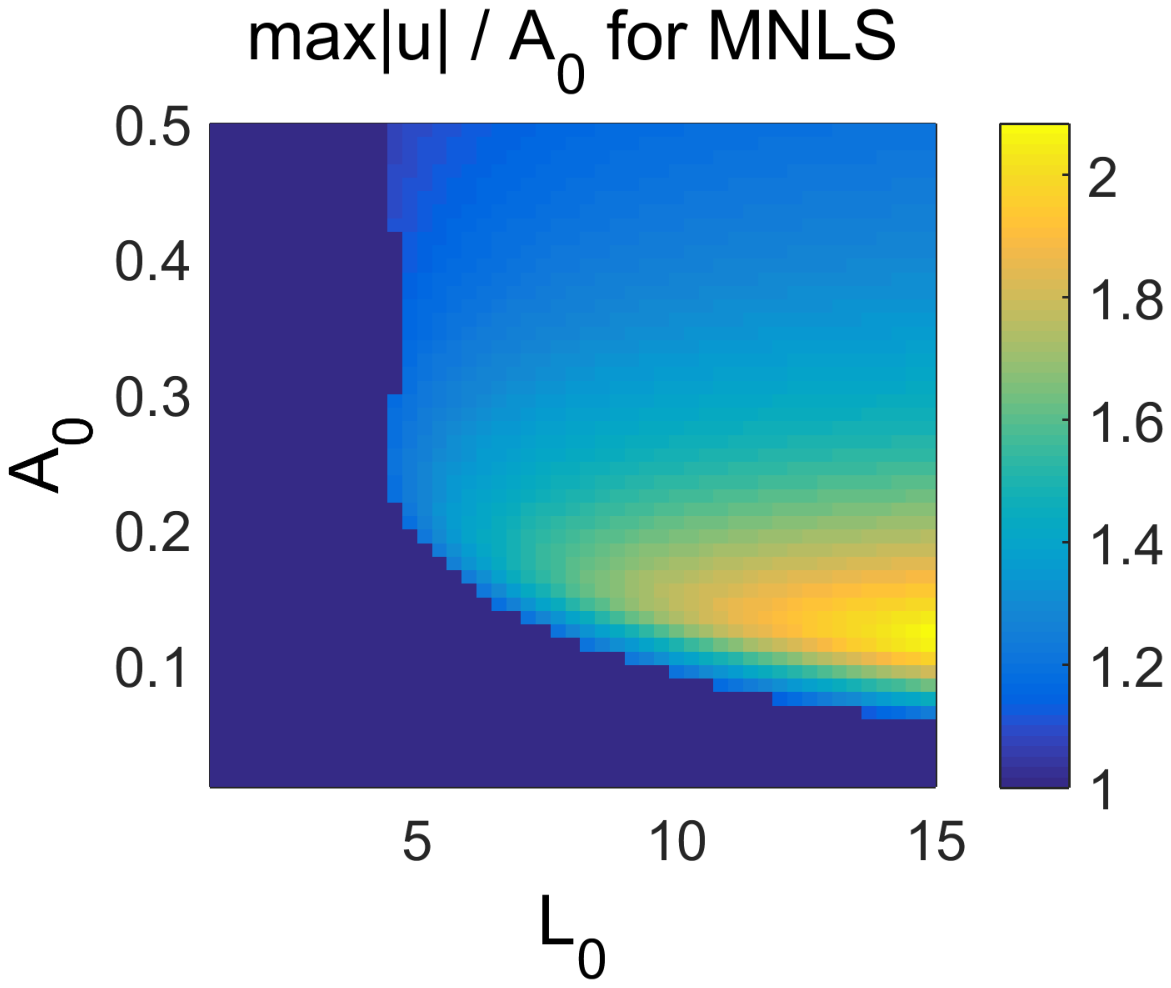}  \\
 \includegraphics[width=0.48\textwidth]{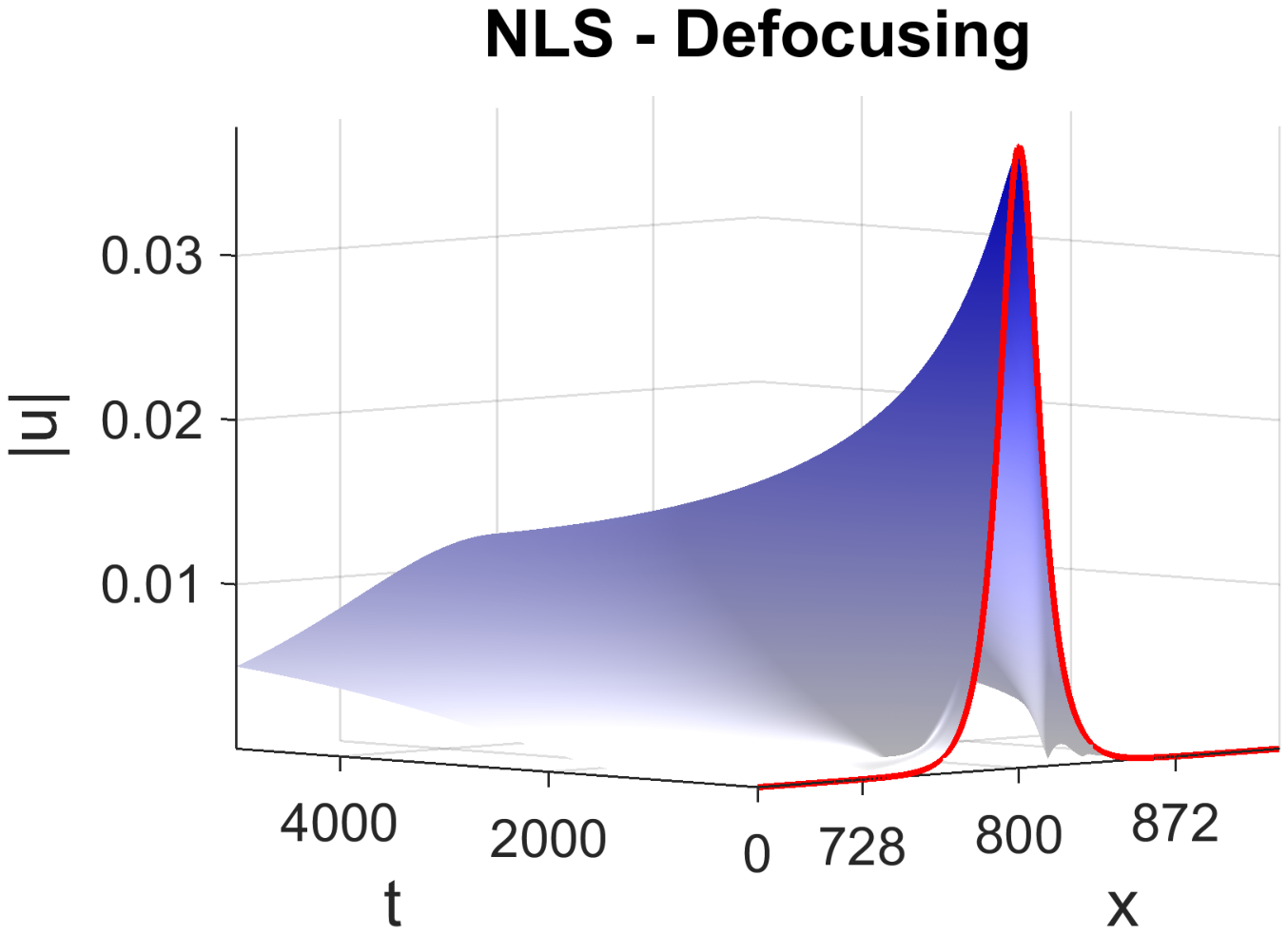}  \includegraphics[width=0.48\textwidth]{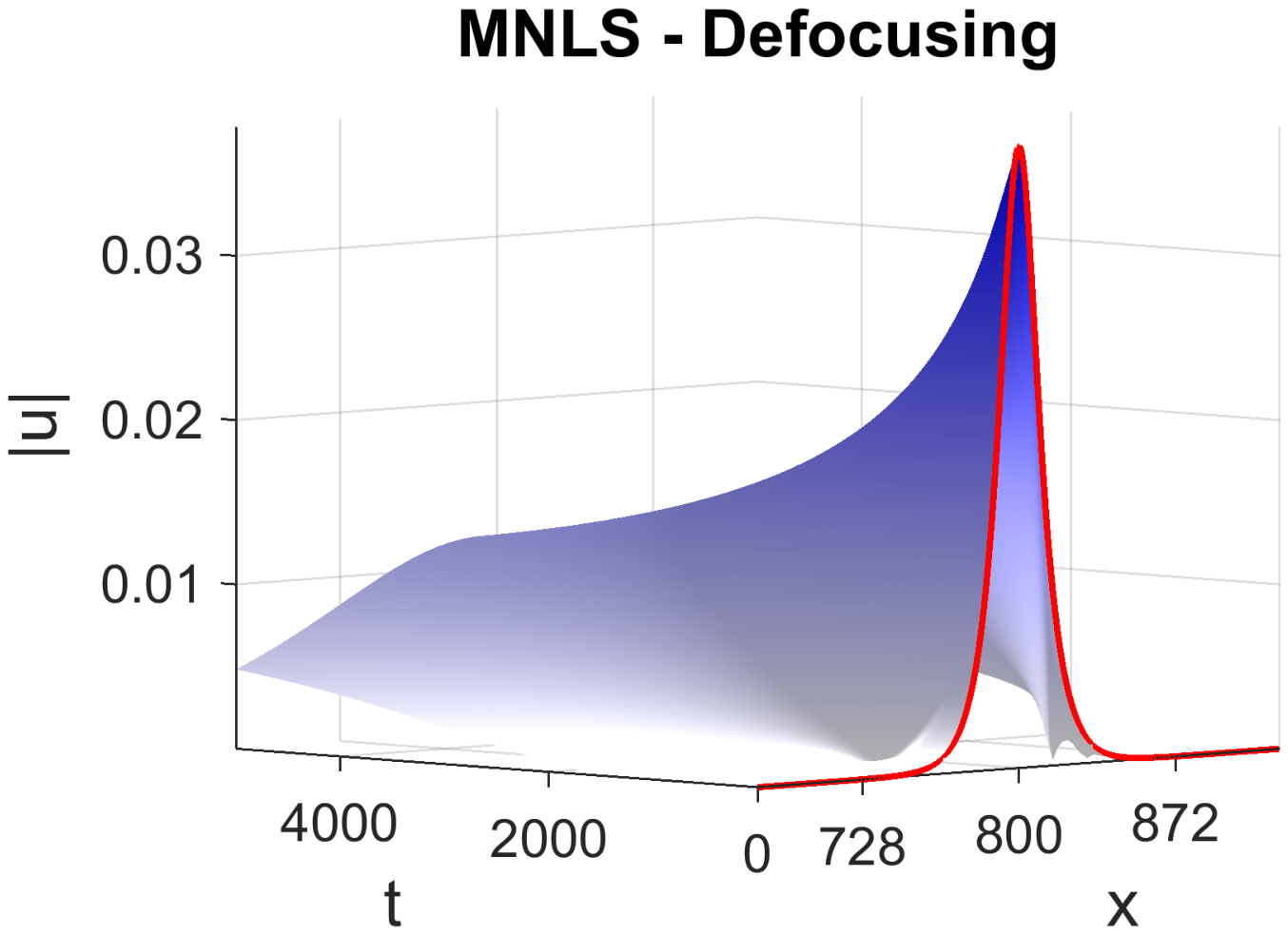} \\
 \includegraphics[width=0.48\textwidth]{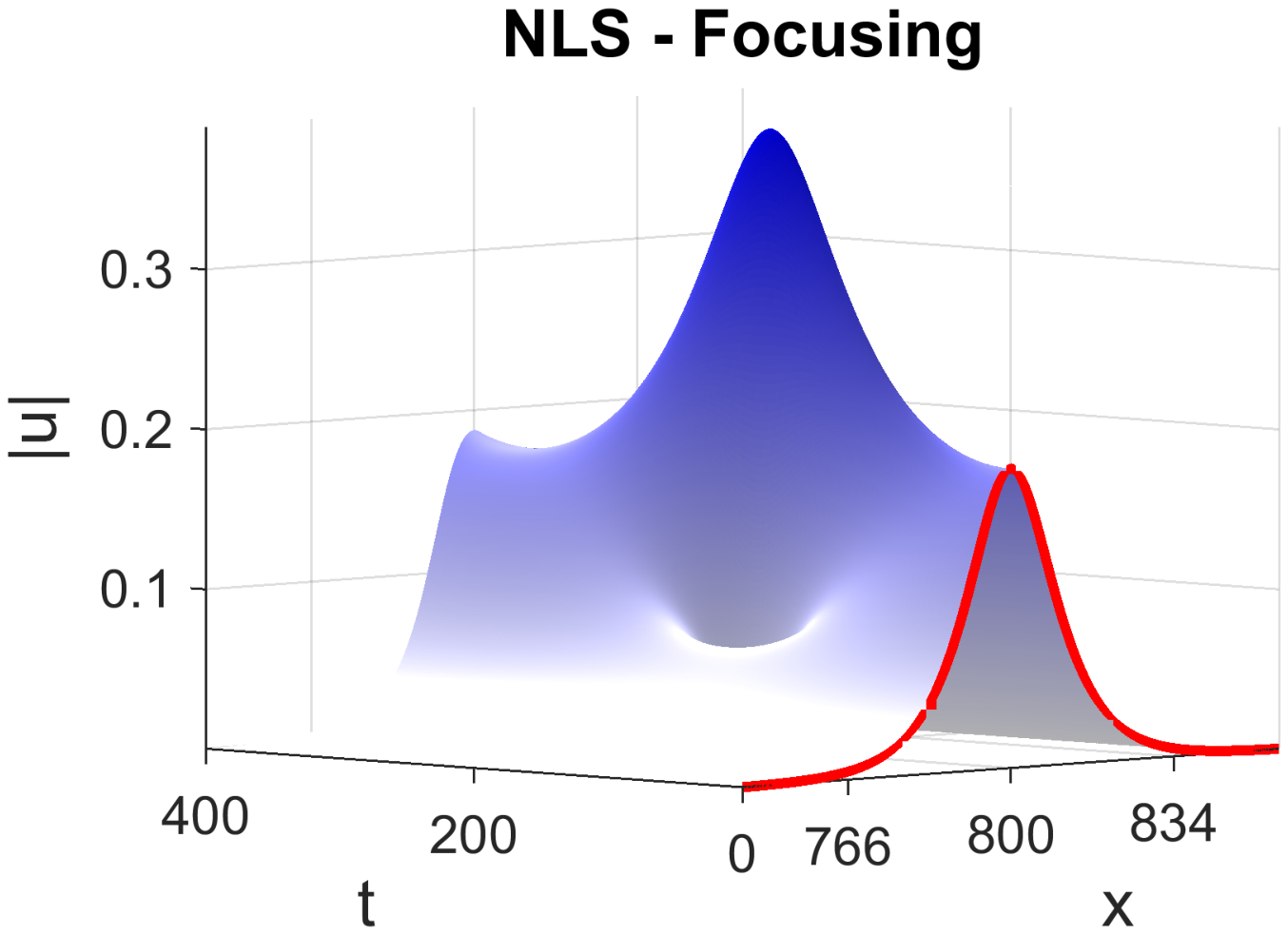} \includegraphics[width=0.48\textwidth]{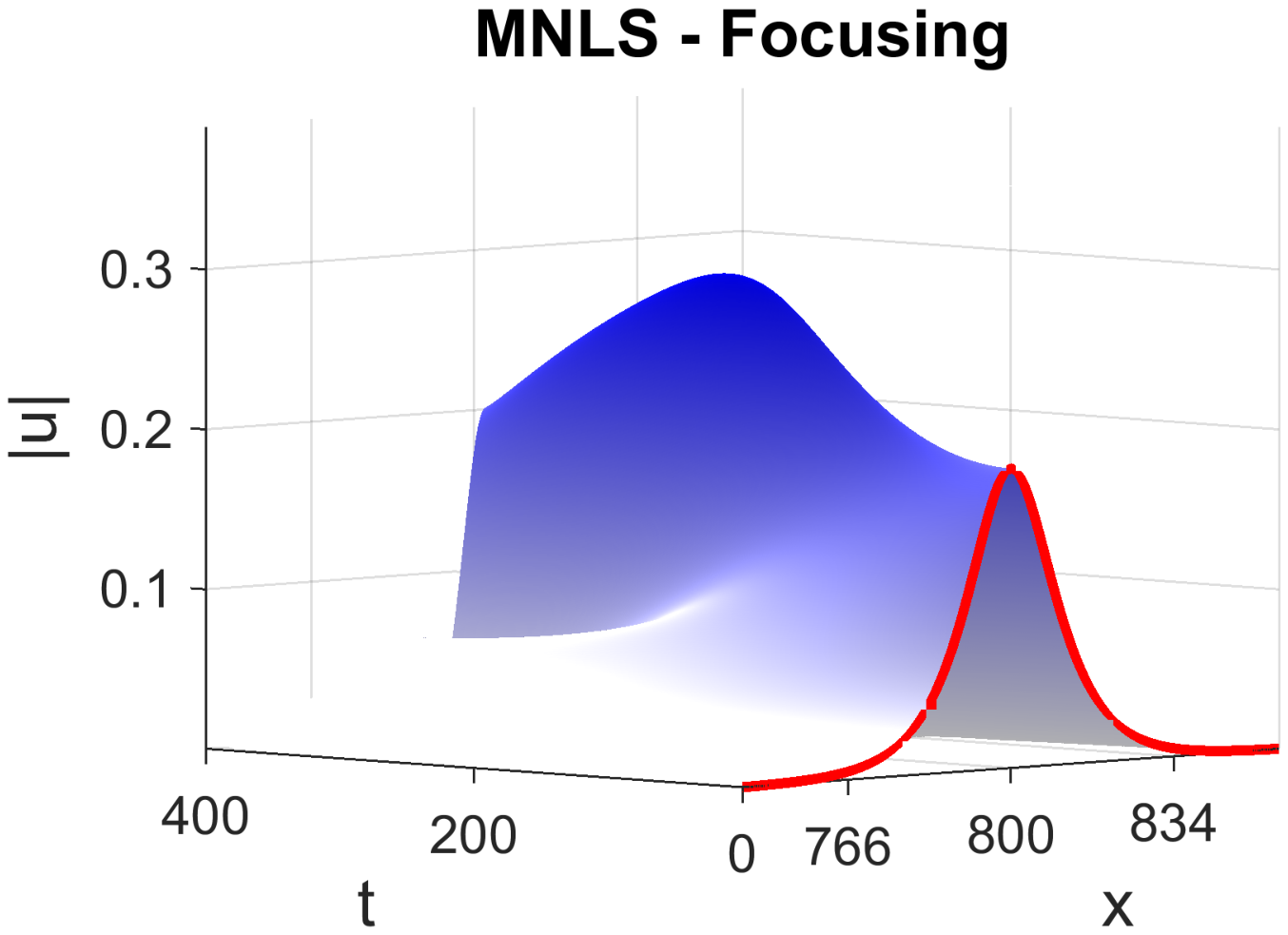} 
	\caption{Top: Amplitude growth factors for localized groups in NLS (left) and MNLS (right).  Middle: examples of defocusing, amplitude decreasing groups for NLS (left) and MNLS (right).  Bottom: examples of focusing, amplitude increasing groups for NLS and MNLS.  The initial conditions for the simulations displayed in the bottom left and bottom right are identical. Although both groups focus, the amplitude grows considerably less in MNLS than in NLS.}
	\label{fig:localID}
\end{figure}
To answer these questions, we numerically evolved hyperbolic secant initial data for many values of $A$ and $L$ for NLS and MNLS.  To measure the degree of focusing a group undergoes, we computed the value of the first spatiotemporal local maximum of $|u|$, and termed this value $u_{max}(A,L)$.  For a defocusing group (middle row, Figure~\ref{fig:localID}), this local maximum will occur at $x=0,t=0$ and we trivially have $u_{max}(A,L) = A$. For a focusing group (bottom row, Figure~\ref{fig:localID}), the group will contract and increase in amplitude.  The group amplitude will eventually reach a maximum and then demodulate, decreasing in amplitude.  To make the focusing behavior clear, in the top two panes of Figure~\ref{fig:localID} we plot the amplitude growth factor $\frac{1}{A} u_{\text{max}}(A,L)$ for NLS (top left pane) and MNLS (top right pane).  This amplitude growth factor describes the degree of focusing which has occurred, with a value of 1 indicating that the group does not grow in amplitude.

We observe that for both NLS and MNLS, there are a range of groups that focus considerably.  However, there are stark differences between group evolution in the two equations owing to the lack of scale invariance in the MNLS equation.  In MNLS, the set of focusing groups is smaller compared with NLS, and many groups that do focus do so to a smaller degree (see example in bottom row, Figure~\ref{fig:localID}).  Particularly, in MNLS there is a smallest focusing length scale where groups thinner than this scale do not focus, regardless of how large their initial amplitude may be.  The scale invariance of NLS, however, precludes such behavior.  A similar lack of focusing behavior at small group length scales was observed by Henderson et. al. in numerical simulations of the fully nonlinear model in one space dimension \cite{henderson1999}.

The nonlinear behavior of focusing groups  can also be described analytically through an adaptive projection of the governing equations on a family of basis elements that respect, by design, certain conservation properties of NLS and MNLS equations. In particular, following \cite{cousinsSapsis2015_PRL} we formulate a reduced-order model that captures the behavior of a localized wave group that has a hyperbolic secant form for as long as this assumption is valid. In addition, the adaptive character of the projection allows for the consideration of the constraints between the amplitude $A$ and length scale $L$ that follow from the conservation properties of NLS and MNLS. This approach gives the following reduced-order model for MNLS \cite{cousinsSapsis2015_PRL}

\begin{align*}
  	\frac{d^2|A|^2}{dt^2} = \frac{K}{|A|^2} \left( \frac{d|A|^2}{dt} \right)^2 - \frac{3|A|^2}{2048L^6}(196|A|^4L^4-64|A|^2L^4+168|A|^2L^2+32L^2+27).
\end{align*}

We stress that the main takeaway from these experiments and reduced order models is the construction of the map $u_{\text{max}}(A,L)$.  That is, given a wave group of amplitude $A$ and length scale $L$, we know the maximally focused amplitude group after its nonlinear evolution via interpolation of our numerical experiments, or approximately using the reduced-order model.  This map will be an essential component of our extreme event predictive scheme, developed in Section~\ref{sec:prediction}.

\section{Probability of Critical Wave Groups in Irregular Wave Fields}

The analysis presented above illustrates the family of unstable wave groups that will focus and eventually lead to an extreme event.  In a particular sea state, dispersion effects create random mixing of different harmonics. To this end, all groups do not occur with equal likelihood--the probability of a particular group occurring is determined by the spectral properties of the field.  For example, in a Gaussian spectrum the spatial field will contain groups unlikely to focus if the energy level $\epsilon$ is small (low amplitude groups) or if the spectral bandwidth $\sigma$ is large (small length scale groups).  Thus, the frequency and nature of the unstable wave groups in a particular field results from the interplay of the nonlinear dynamical properties of the system (determined by the underlying physics) and the statistical properties of the background field (captured by the spectrum), where dispersion is the dominant mechanism.

\begin{figure}[H]
	\centering
	\includegraphics[width=0.9\textwidth]{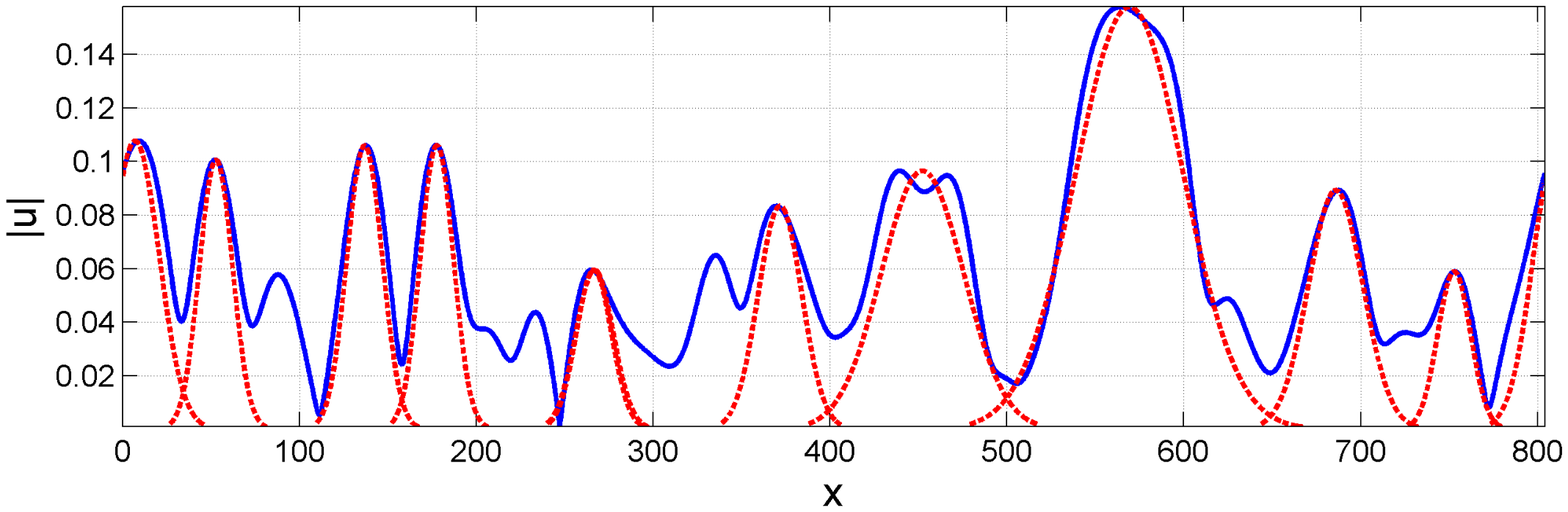} \\
	\includegraphics[width=0.4\textwidth]{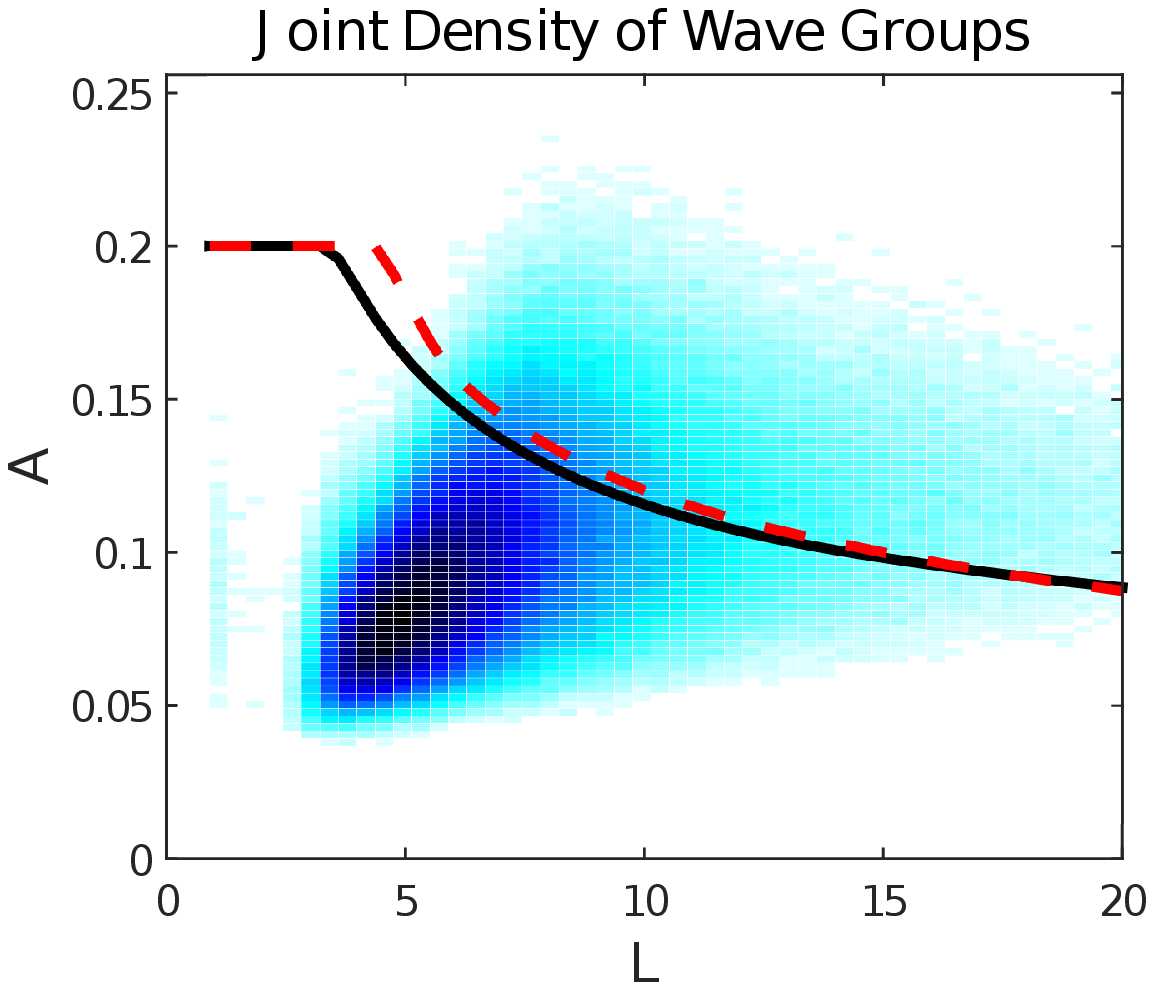} \includegraphics[width=0.4\textwidth]{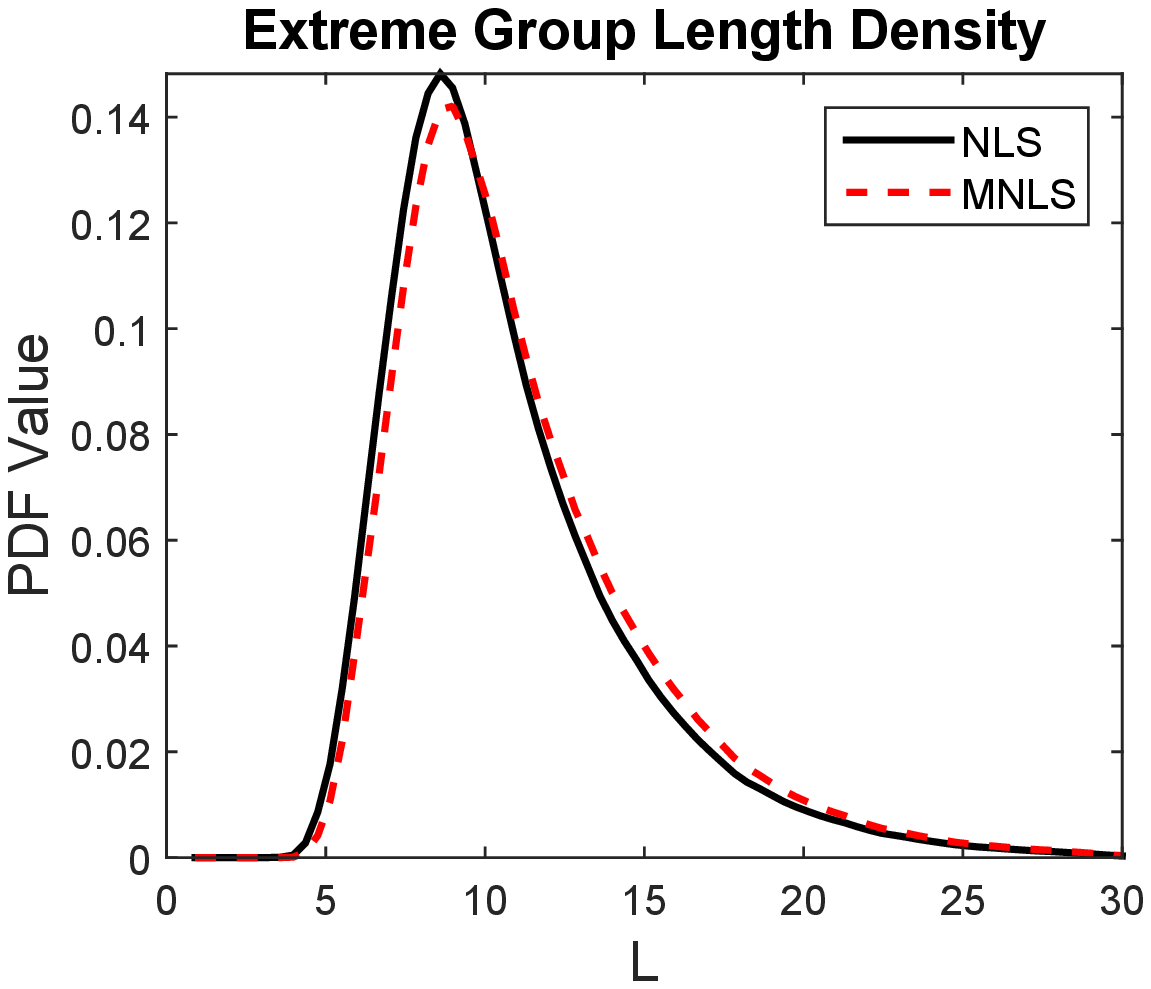} \\
	\includegraphics[width=0.4\textwidth]{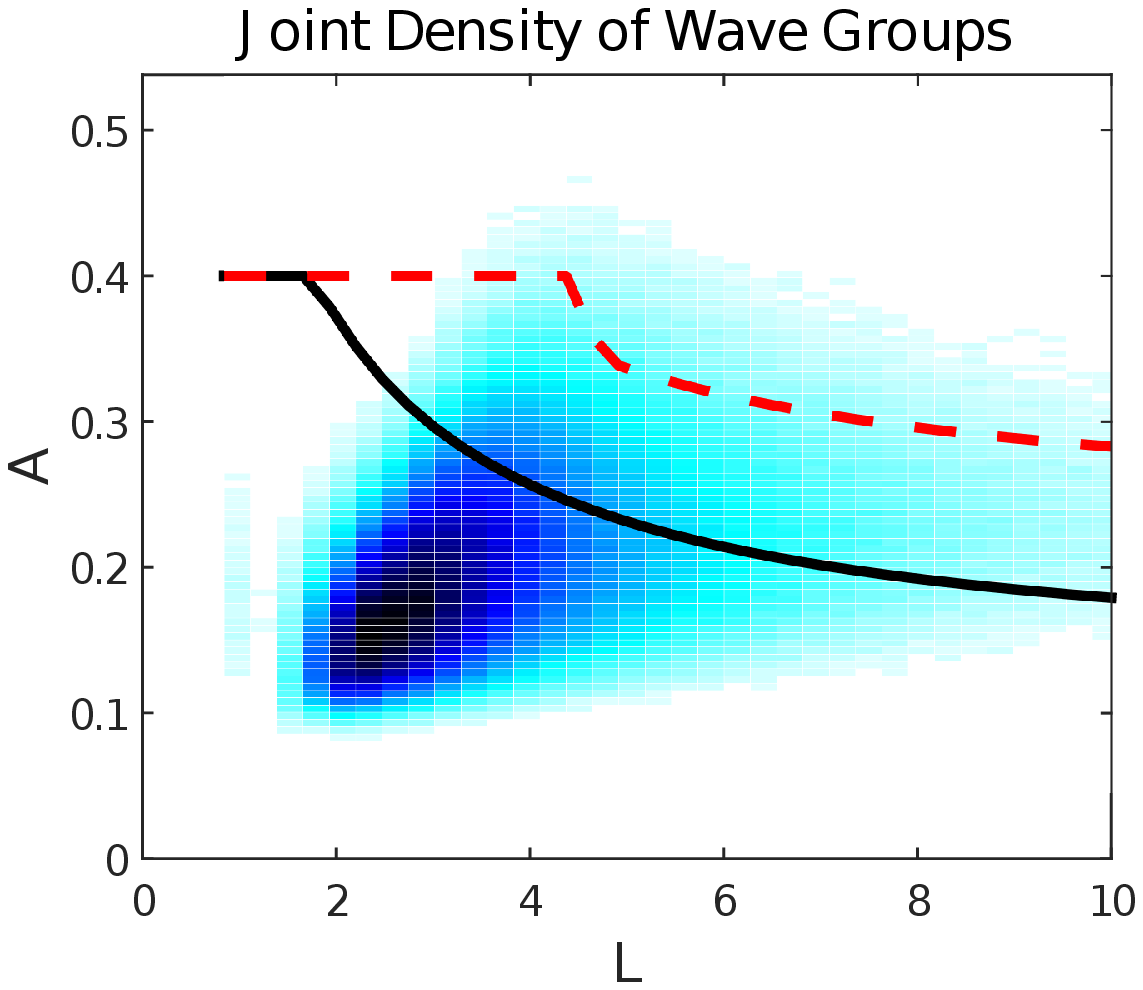} \includegraphics[width=0.4\textwidth]{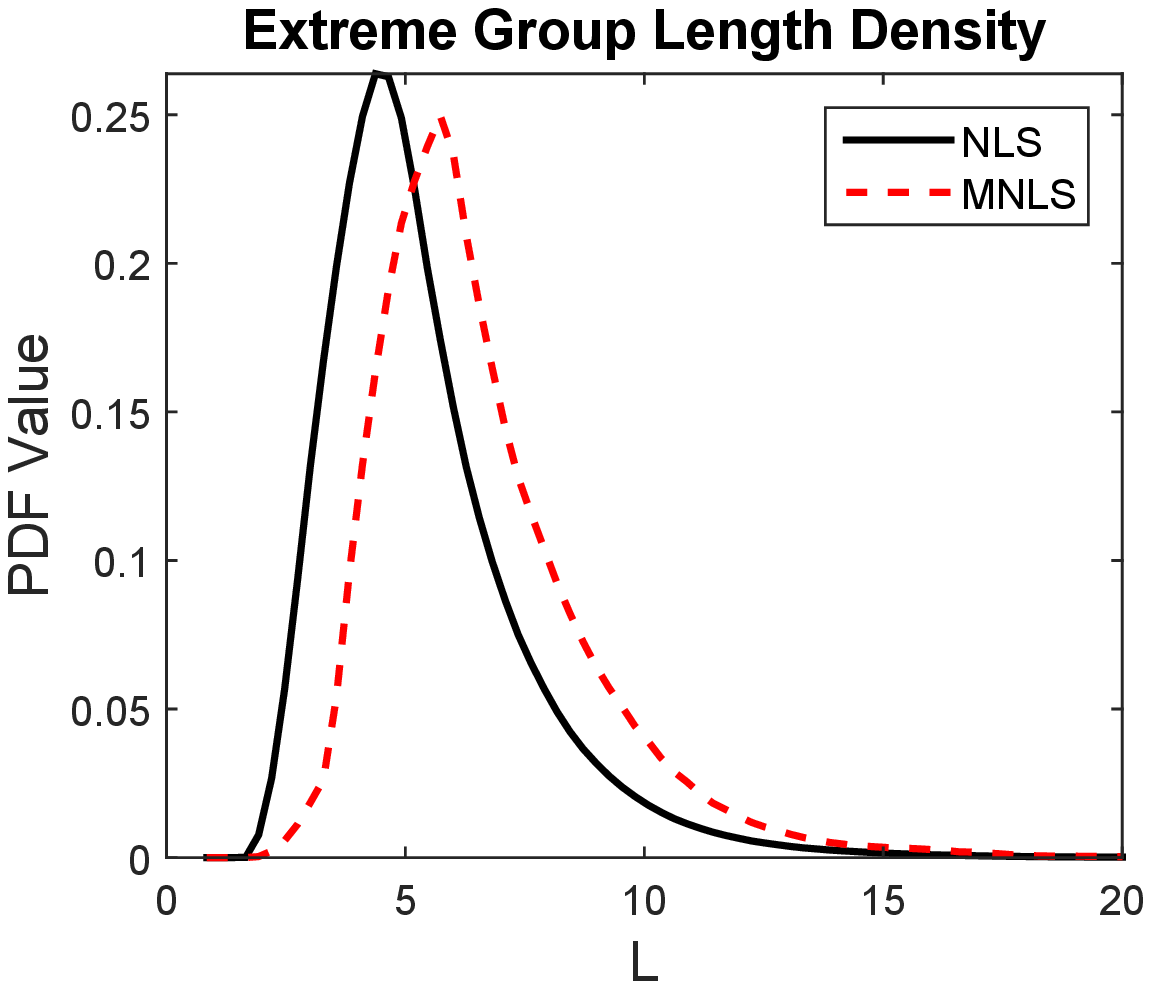}
	\caption{Top: Dominant wave groups (red, dashed) determined by applying the group detection algorithm to the field $|u(x)|$ (blue, solid). Middle Row: Joint density of group amplitude and length scale (left) and length scale density for extreme triggering groups (right), BFI = 1.4, $\epsilon = 0.05, \sigma = 0.1$. Bottom row: idem for BFI = 1.4, $\epsilon=0.1,\sigma=0.2$.  In each joint density plot we overlay the curve that separates groups that would focus (if isolated) to form extreme events, NLS curve is black, solid, MNLS curve is red, dashed.}
	\label{fig:groupStatistics}
\end{figure}

We now describe a procedure to directly quantify how the statistics of the background field interact with the underlying nonlinear dynamics to create extreme events.  To do so, for a variety of Gaussian spectra we compute the joint density of group amplitude and length scale. To compute this density, we use a scale selection algorithm to identify groups as well as their associated length scales and amplitudes (see Appendix~\ref{app:scaleSelection} for details on the scale selection algorithm).  In the top pane of Figure~\ref{fig:groupStatistics}, we display a particular field $|u(x)|$ as well as the groups identified by the group detection algorithm, showing that this algorithm appropriately picks out the dominant groups. We then compute the joint probability density of group amplitude and length scale by applying this group identification algorithm to 50,000 realizations of Gaussian spectrum random fields.

For each length scale $L$, we determine the smallest group amplitude required to trigger an extreme event.  That is, if the extreme event threshold is $H_E = 4\epsilon$, we find, for each $L$, the smallest $A$ such that $u_{\text{max}}(A,L) \geq H_E$.  Denoting this amplitude as $A^*$, we trivially have $A^* \leq H_E$.  This procedure describes a curve $A^*(L)$, where isolated groups located above this curve in the $(L,A)$ plane would yield an extreme event, and those below this curve would not.  We overlay this curve over the joint densities of group amplitude and length scale displayed in Figure~\ref{fig:groupStatistics}.

For a given spectrum, we may determine the frequency and nature of extreme event-triggering groups (i.e. those lying above the respective curves in Figure~\ref{fig:groupStatistics}.  As expected, increasing the energy level or decreasing the spectral bandwidth increases the number of these extreme-triggering groups.  This analysis provides a concrete, wave-group based explanation of the development of heavy tails via nonlinear interactions in high BFI regimes.  However, unlike traditional BFI-based analysis, this wave group analysis directly incorporates the lack of scale invariance in MNLS. This fact can be observed in Figure ~\ref{fig:groupStatistics} (middle and bottom rows) where we display $(A,L)$ densities for two different spectra with the same BFI. We note the scale invariance property of NLS (note that the two black curves are identical, albeit the axes rescale) and how this contrasts to the corresponding curves (red dashed lines) for the MNLS, which change between the two spectra even though the BFI index remains the same. 

This statistical instability analysis allows us to characterize the properties of extreme event-triggering groups in a maximum likelihood sense. Specifically, for a given spectrum, we can compute the density of the length scale of groups that would generate extreme events (i.e. those lying above the curves in the left column of Figure~\ref{fig:groupStatistics}). In Figure ~\ref{fig:groupStatistics} (right column) we also display examples of these length scale densities for the two different spectra. The concentration of these densities around their peak value suggests that for a given spectrum there is a most likely extreme event triggering length scale, $L_E$.  As the distribution is fairly narrow, we expect that the majority of extreme events that occur will be triggered by localization at length scales close to $L_E$.  In Section~\ref{sec:prediction} we will use this fact to develop a predictive scheme based on projecting the field onto an appropriately tuned set of Gabor wavelets.

\section{Prediction of Extreme Events}
\label{sec:prediction}

In this section, we describe the central result of this paper, schemes for prediction of extreme events.  Our goal is to develop schemes that provide reasonably precise spatiotemporal predictions of upcoming extreme events.  These schemes also must require less computational cost then solving the full envelope equations.

We describe two predictive schemes that meet these criteria.  First, we develop an algorithm to predict extreme events by identifying the dominant wave groups in a given field, and use our results from Section~\ref{sec:local} regarding evolution of localized groups to predict whether this group will trigger an extreme event.  Second, we develop a scheme based on projecting the field onto a carefully tuned set of Gabor modes.  We find that large values of a certain Gabor coefficient indicate that an upcoming extreme event is likely.  This Gabor-based scheme is nearly as reliable as the group detection scheme, yet requires a remarkably small computational cost.

\subsection{Prediction by Wave Group Identification}
\label{sec:multiscale}

Here we describe a straightforward scheme for advance prediction of extreme events via wave group identification.  For a given field, we apply the group identification algorithm described in Appendix~\ref{app:scaleSelection} to the envelope.  This gives the spatial location, amplitude $A$, and length scale $L$ of each wave group.  We then predict the future focused amplitude of the group by evaluating $u_{max}(A,L)$, where $u_{max}$ is the numerically constructed function from our prior study of localized groups in Section~\ref{sec:local}. If $u_{max}(A,L)$ is at least 95\% of the extreme event threshold $H_E$, then we predict that an extreme event will occur.  We choose this conservative prediction threshold in order to minimize the number of false negatives (extreme events that we fail to predict).

\begin{figure}[H]
	\centering
	\includegraphics[height=2in,width=\textwidth]{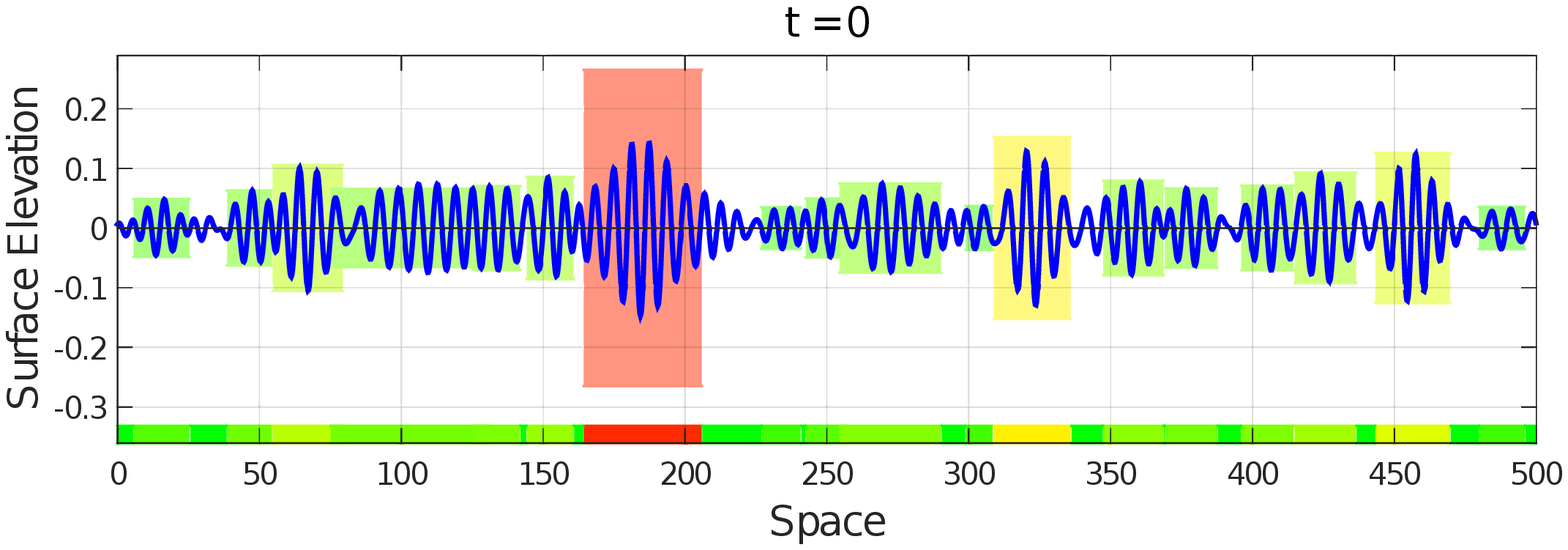} \\ [1em]
	\includegraphics[height=2in,width=\textwidth]{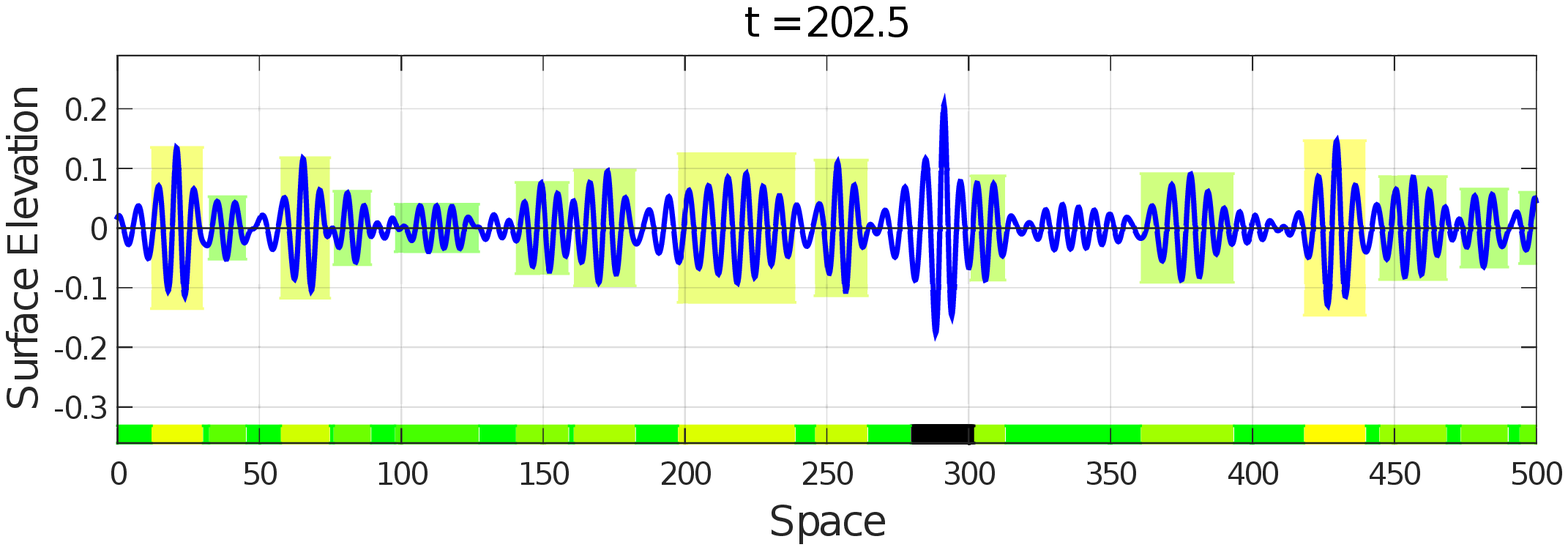} \\ [1em]
	\includegraphics[height=2in,width=\textwidth]{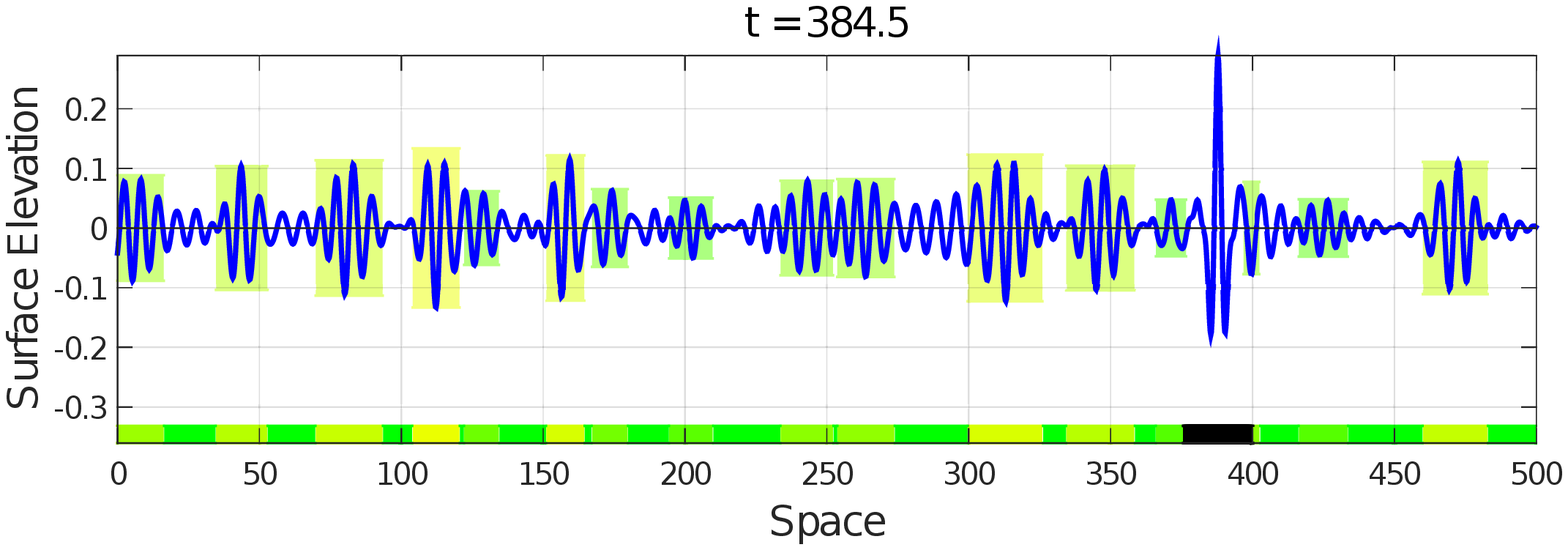}
	\caption{Top: Initial conditions for a simulation of MNLS.  Our scheme identifies a group around $x=190$ which we predict will grow to form a large extreme event.  Middle: Group in initial stages of focusing, breaks the extreme event threshold $H_E=0.2$ near $x=290$. Bottom: group is fully focused and attains its maximum amplitude near $x=390$.}
	\label{fig:multiscaleFS}
\end{figure}

In Figure~\ref{fig:multiscaleFS}, we display an example output of our predictive scheme for a simulation of MNLS with initial conditions generated via a Gaussian spectrum with random phases ($\epsilon = 0.05, \sigma = 0.1$, BFI = 1.4, $H_E$=0.2). We display the spatial dependence of the surface elevation for three different values of time.  In this simulation, the surface elevation first exceeds $H_E$ around $t=200$ near $x=300$. After exceeding this threshold, the extreme event continues to focus, eventually reaching a maximum of approximately 0.3 at $t\approx 385, x\approx 390$ (Figure~\ref{fig:multiscaleFS}, bottom pane).

We highlight each wave group with a rectangle whose height is equal to the predicted focused amplitude of the group, with a red colored rectangle indicating that we predict that the group will focus to form an extreme event.  We see that our scheme identifies the group that will trigger the extreme event far in advance.  The initial prediction occurs at the beginning of the simulation at $t=0$, 200 time units ($\approx$32 temporal wave periods) before the elevation crosses the extreme event threshold $H_E=0.2$, and nearly 400 time units before the extreme event reaches its maximal amplitude.  Perhaps most importantly, the prediction occurs while the elevation is at the relatively modest value of 0.147.

We emphasize that our scheme accurately gives the spatiotemporal location where the extreme event will occur.  The group that we predict will focus to an extreme event (red rectangle in Figure~\ref{fig:multiscaleFS}, top pane), has a length scale of 10.3 and amplitude 0.147.  In MNLS, a localized hyperbolic secant initial profile with these characteristics will focus to a maximum amplitude of 0.264 after 351 time units, meaning that our scheme predicts a wave of this amplitude at $t=351, x=361$.  To predict this spatial location we use the linear group velocity and the fact that the group is located at $x=185$ initially.  This agrees well with the observed dynamics of the simulation of the full field using MNLS--the identified group reaches an actual maximum of 0.289.

To test the reliability of this scheme, we implemented it on 100 simulations each of NLS and MNLS with BFI = 1.4 (see Appendix~\ref{app:scaleSelection} for details on these simulations). Here we only discuss the MNLS results, as the NLS results are similar and MNLS is the more physically relevant equation.  In these 100 simulations, there were 336 extreme events.  We predicted all of these extreme events in advance--there were no false negatives.  There were 91 instances where we predicted an extreme event but one did not occur, giving a false positive rate of 21.3\%.  For our correct predictions, the average warning time (the amount of time before the prediction began and the onset of the extreme event) was 153 time units ($\approx$24 temporal wave periods).

\begin{figure}[H]
	\centering
	\raisebox{0.1\height}{\includegraphics[width=0.375\textwidth]{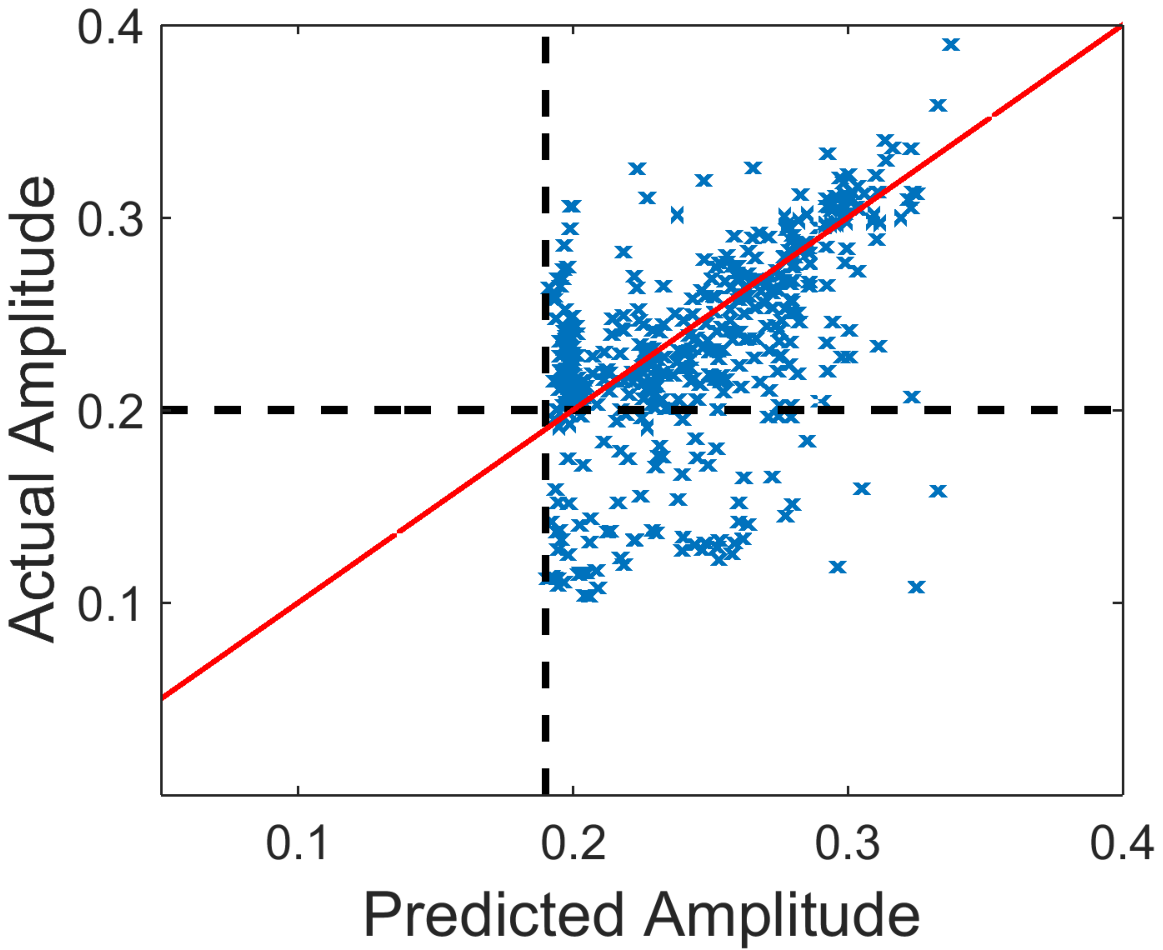}} \includegraphics[width=0.615\textwidth]{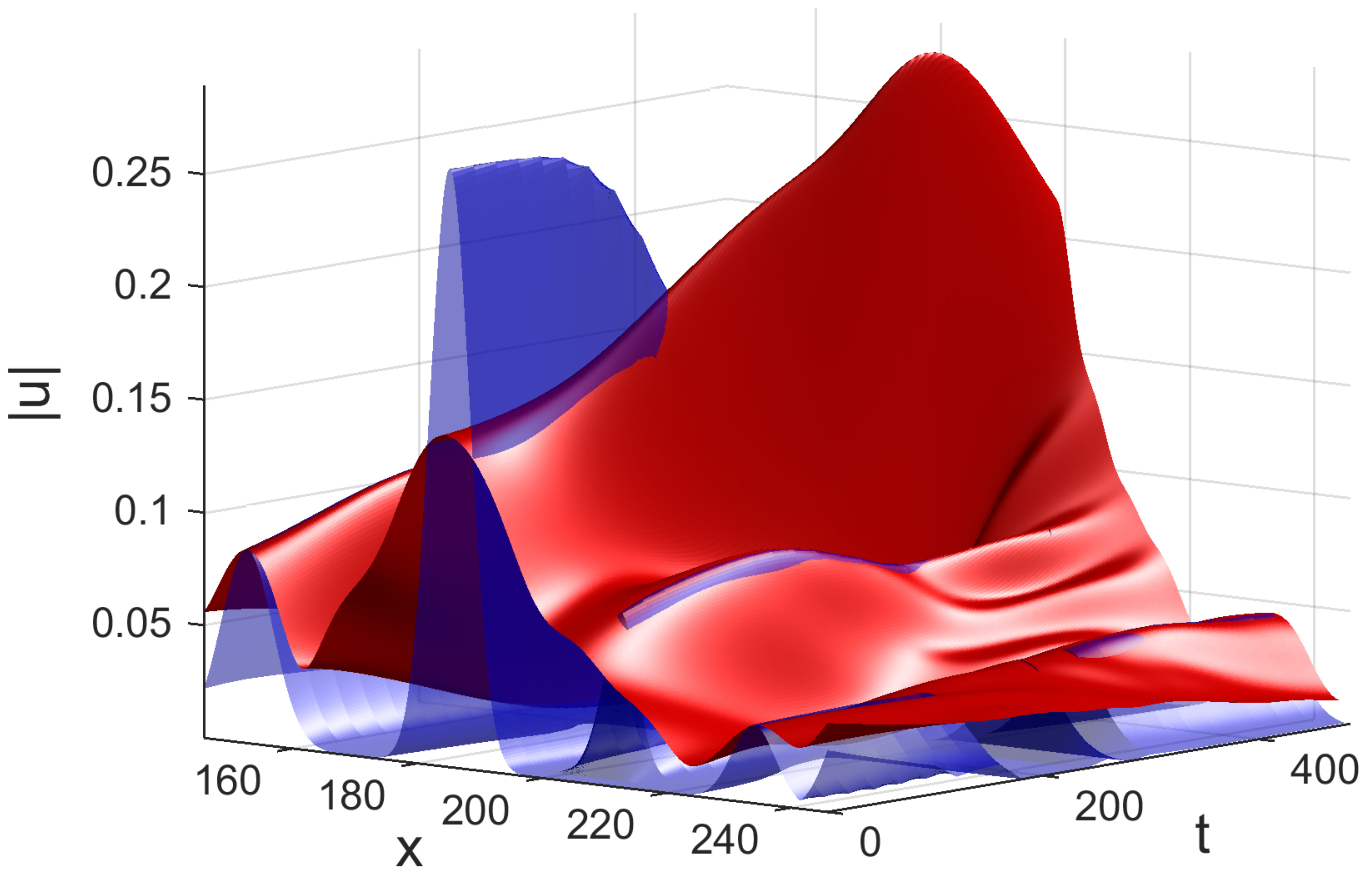}
	\caption{Left: scatterplot of predicted/actual amplitudes as well as the line predicted=actual (red). Note that the vertical dashed line is located at $0.95H_E$ to reduce false negatives (as discussed in the text). Right: spatiotemporal dependence of $|u|$ (red) and predicted future amplitude (blue).}
	\label{fig:multiscaleContinuous}
\end{figure}

Our scheme has value beyond a binary predictor of extreme events.  In the left pane of Figure~\ref{fig:multiscaleContinuous}, we display a scatterplot showing the relationship between the predicted fugure amplitude and the actual future amplitude of the wave field.  We observe that our predictor reliably estimates the future amplitude in a continuous sense.  In particular, in addition to predicting when an upcoming extreme event is likely, our scheme predicts when a particularly \emph{large} extreme event is upcoming.

To further illustrate the skill of our scheme, in the right pane of Figure~\ref{fig:multiscaleContinuous}, we plot the surface $|u(x,t)|$ in red, as well as the predicted future group shape in blue.  The field displayed here is the same field displayed in Figure~\ref{fig:multiscaleFS} in a coordinate frame moving with the linear group velocity.   The surface plots in Figure~\ref{fig:multiscaleContinuous} provide a visualization of the skill of our scheme in predicting the future amplitude of the future extreme wave, as well as the spatial location at which it will occur.  The reason the blue surface near the extreme event decays and vanishes around $t=200$ is because we locally turn off the predictor while an extreme event is occurring ($|u|>H_E$).

\subsection{Prediction by Gabor Projection}
\label{sec:gabor}

We now describe an alternative reliable prediction scheme that requires negligible computational cost.  In this scheme, we predict upcoming extreme events by projecting the field onto a set of carefully tuned Gabor modes. This approach is similar in spirit to our extreme event predictive scheme for the model of Majda, McLaughlin, and Tabak (MMT) \cite{cousinsSapsis2014}.  This projection requires only a single convolution integral, so its computation is extremely cheap.  Even at this low cost, this scheme reliably predicts upcoming extreme events with spatiotemporal skill.

As we showed in Section~\ref{sec:local}, for a given spectrum we can compute the joint density of wave group amplitude and length scale.  Using our study of isolated localized groups, we can then compute the conditional density of wave group properties for groups that will trigger extreme events.  This gives, among other things, the density of group length scales for groups that would focus to form an extreme event (refer to Figure~\ref{fig:groupStatistics}, right column).  From this, we can compute the spatial length scale $L_G$ with the maximum likelihood of triggering an extreme event.  Due to the narrowness of the distribution of extreme event-triggering group length scales, we expect that extreme events will be preceded by energy localization at a length scale close to $L_G$.

To predict extreme events, we estimate the energy concentrated in scale $L_G$.  To do so, we ``project" the field onto the set of Gabor basis functions $v_n(x;x_c)$, which are complex exponentials multiplied by a Gaussian window function.  Projecting the envelope onto these Gabor functions gives the the Gabor coefficients $Y_n(x_c,t)$:

\begin{align*}
	v_n(x;x_c) &= e^{i\pi n (x - xC) /L_G} \exp\left[ -\frac{(x-x_C)^2}{2L_G^2} \right], \\[1em]
	Y_n(x_c,t) &= \langle u(x,t),v_n(x;x_c) \rangle / \langle v_n(x;x_c), v_n(x;x_c) \rangle.
\end{align*}
We expect that a large value of $Y_0$ at a spatial point $x_c$ indicates that an extreme event is likely in the future near $x_c$ in space (in a frame moving with the group velocity).  To confirm this, we compute the following family of conditional distributions.

\begin{align}
F_{\mathcal{Y}_{0}}(\mathcal{U})\triangleq\mathcal{P}\left(\max_{\substack{|x^{*}-x_{c}|<L_G\\
t^{\ast}\in\lbrack t+t_A,t+t_B]
}
}|u(x^{\ast},t^{\ast})|>\mathcal{U}\text{ \ }\Big|\text{ }|Y_{0}(x_{c},t)|=\mathcal{Y}_{0}\right).\label{eq:Y0CondPDF}
\end{align}
That is, given a particular current value of $Y_0$, we examine what are the statistics of the envelope $u$ in the future. Here we choose $t_A = 50$ and $t_B = 350$ from the time required for a group of length scale of $L_G$ to focus to form an extreme event.  We compute the statistics (\ref{eq:Y0CondPDF}) from 200 simulations of NLS/MNLS with Gaussian spectra and random phases with $\epsilon = 0.05, \sigma =0.1$, BFI = 1.4.

\begin{figure}[H]
	\centering
	\includegraphics[width=\textwidth]{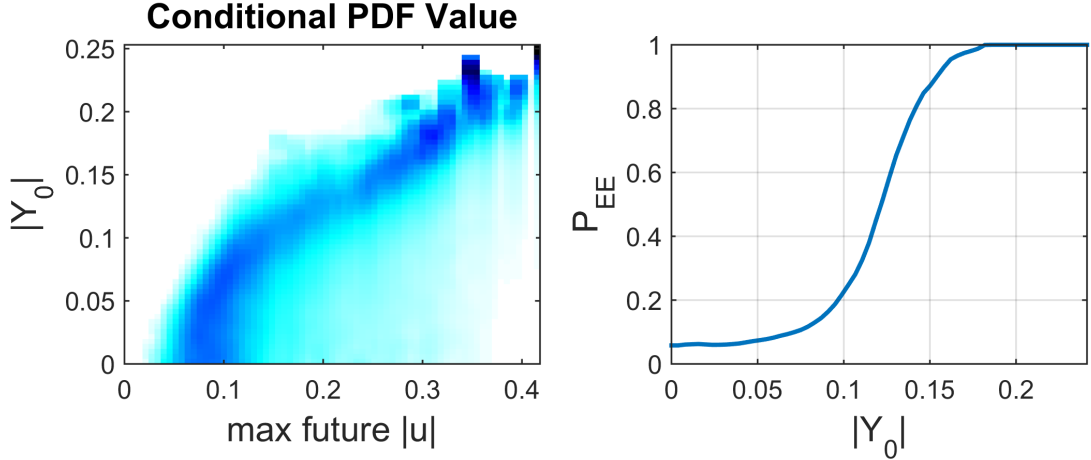}
	\caption{Left: Family of conditional densities of current $|Y_0|$ and future $|u|$, corresponding to (\ref{eq:Y0CondPDF}). Right: probability of upcoming extreme event as a function of $|Y_0|$}
	\label{fig:condY0}
\end{figure}

In the left pane of Figure~\ref{fig:condY0}, we display the family of conditional densities of future $|u|$ for a range of values of $|Y_0|$.  These densities show that when $|Y_0|$ is large, $|u|$ is likely to be large in the future.  From these conditional statistics, we compute the probability of an upcoming extreme event $P_{EE}$ as a function of current $Y_0$ by integrating over $(H_E,\infty)$.  This function is displayed in the right panel of Figure~\ref{fig:condY0}.  $P_{EE}$ has a sigmoidal dependence on $Y_0$: if $Y_0$ is large enough than an upcoming extreme event is nearly guaranteed, while if $Y_0$ is small enough than an upcoming extreme event is highly unlikely.

\begin{figure}[H]
	\centering
	\includegraphics[width=\textwidth]{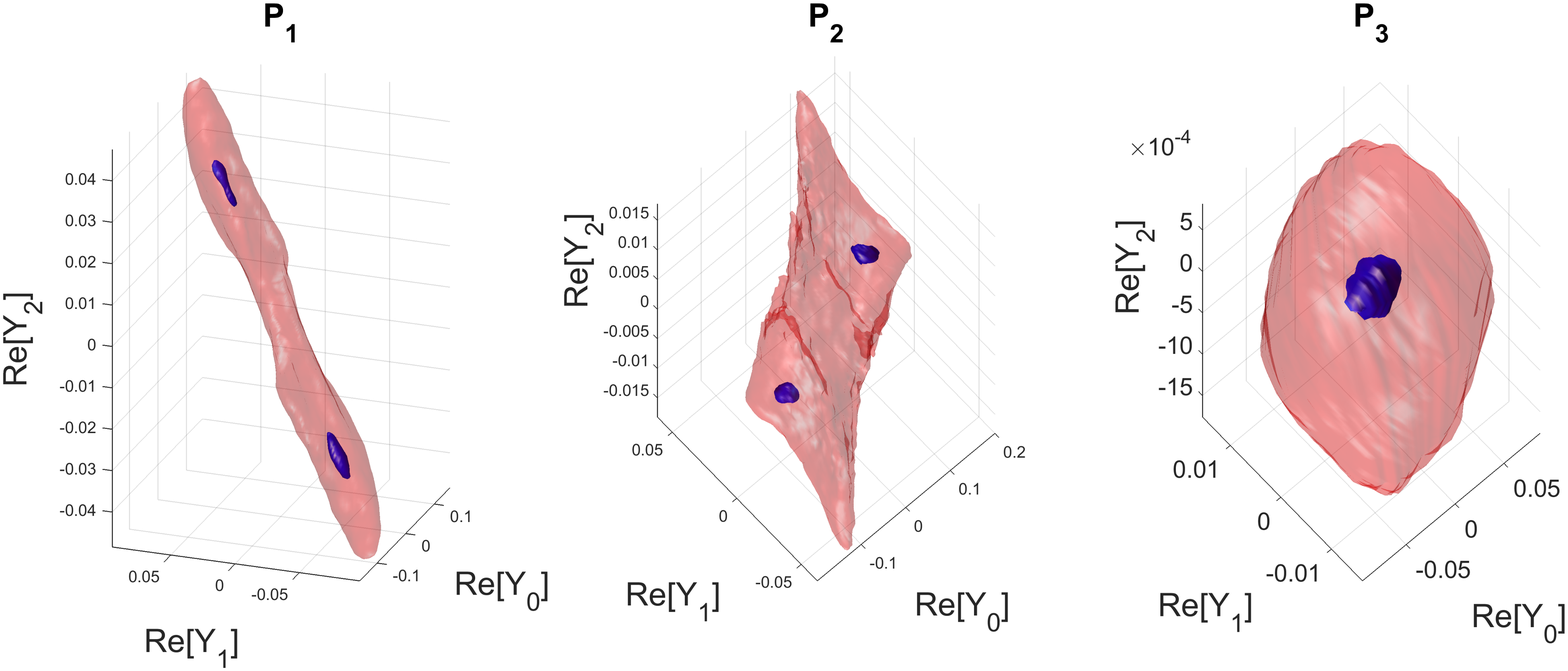} \\ [1em]
	\includegraphics[width=0.3\textwidth]{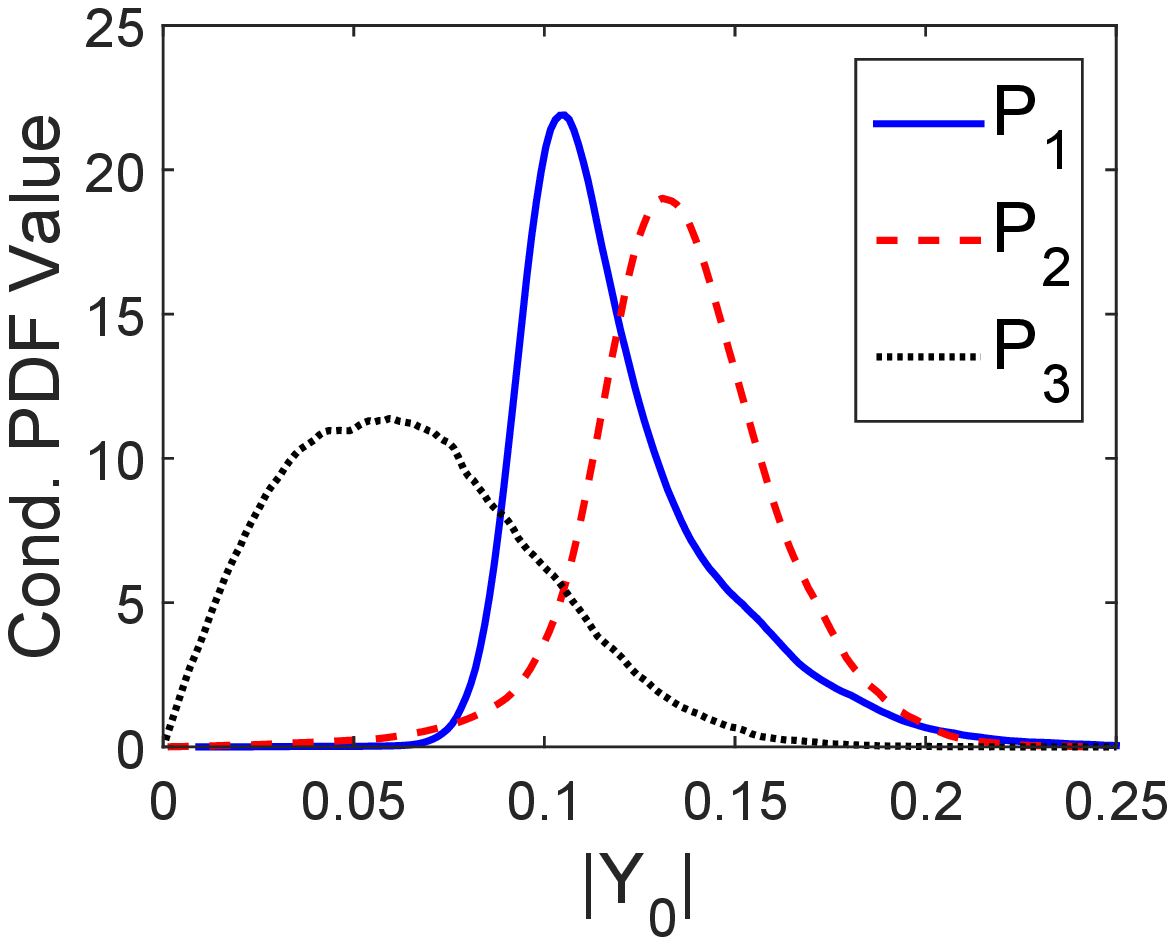} \includegraphics[width=0.3\textwidth]{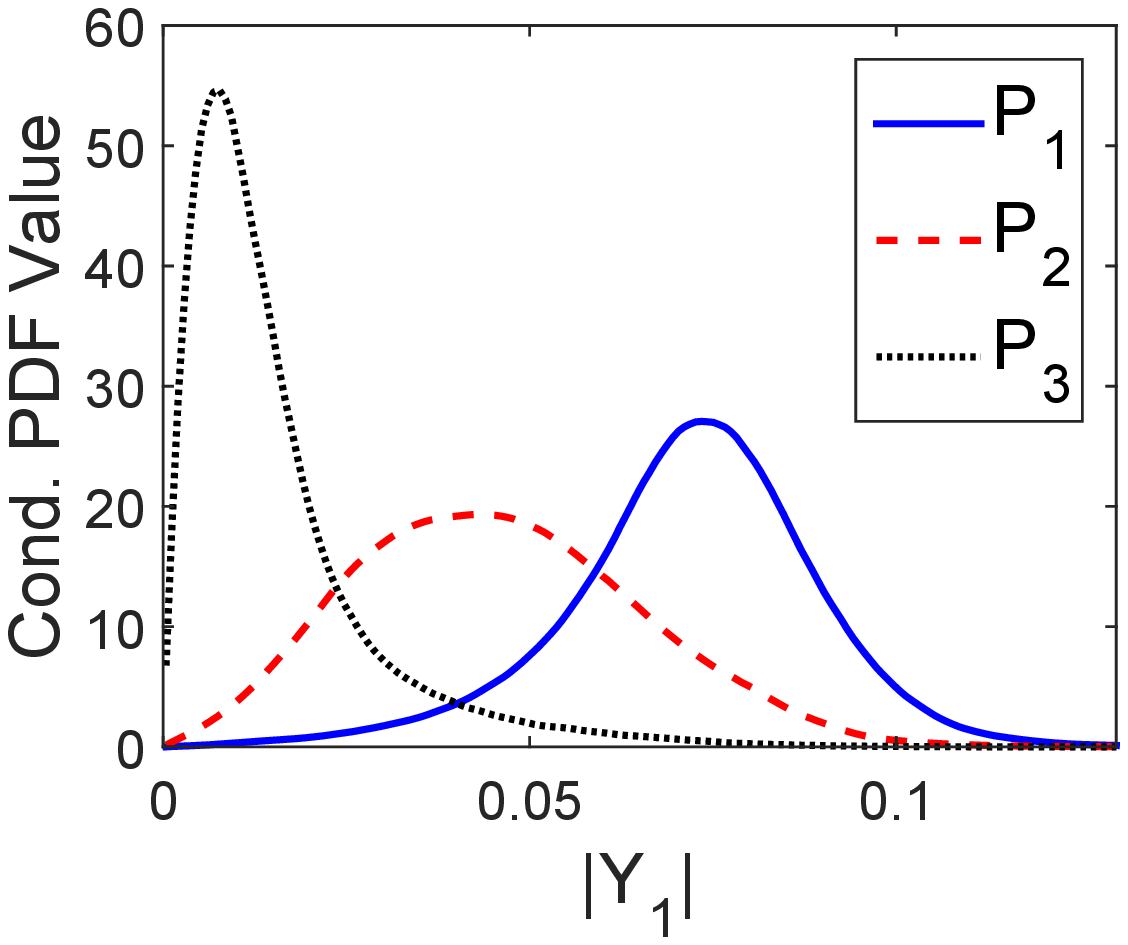} \includegraphics[width=0.3\textwidth]{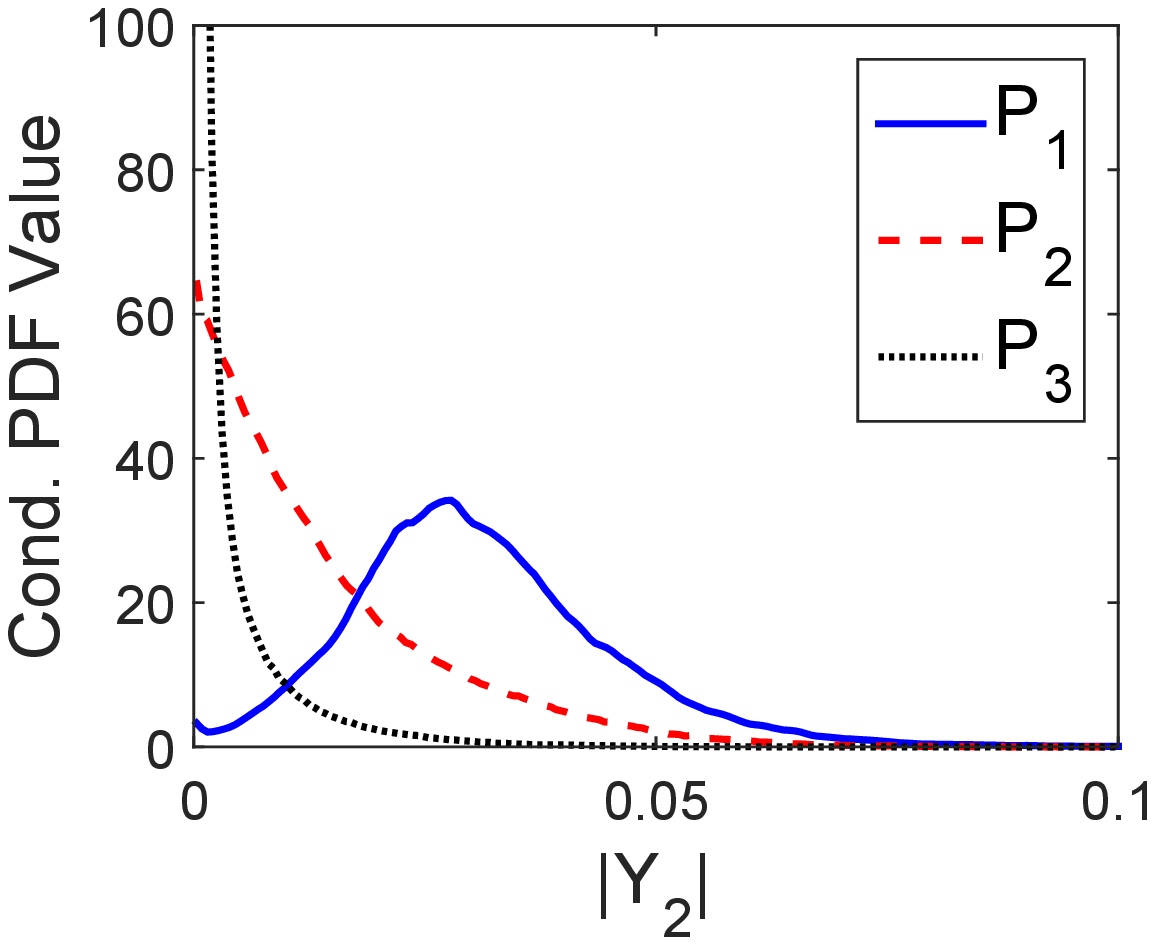}
	\caption{Top: Level surfaces of the joint density of the real parts of $Y_0, Y_1,$ and $Y_2$ during ($P_1$, left), before ($P_2$, middle) and far from ($P_3$, right) an extreme event.  Bottom: marginal statistics of $Y_0$ (left), $Y_1$ (middle), and $Y_2$ (right) during ($P_1$), before ($P_2$), and far from ($P_3$) an extreme event.}
	\label{fig:gaborJointStats}
\end{figure}

We emphasize that $Y_0$ becomes large distinctly before the extreme event occurs. The conditional statistics shown in Figure~\ref{fig:condY0} pair a value of $Y_0$ with a maximum value of $|u|$ which occurs at least $t_A=50$ time units in the future.  To further illustrate this point, we statistically investigate the energy exchanges between the various Gabor modes.  To do so, we compute the statistics of the Gabor coefficients during, before, and far from an extreme event. We display these Gabor statistics in Figure~\ref{fig:gaborJointStats}.  We see that away from extreme events, we have nearly Gaussian statistics for the coefficients.  In this regime the coefficients also appear to be uncorrelated. Before an extreme event, the coefficients are larger--$Y_0$ in particular is quite large.  During an extreme event, $Y_0$ is on average \emph{smaller} than before extremes (see Figure~\ref{fig:gaborJointStats}, bottom left).  In the formation of the extreme events, $Y_0$ decreases by transferring energy to $Y_1$, which is largest during the extreme events.  During the extreme events, the Gabor coefficients are also strongly correlated (Figure~\ref{fig:gaborJointStats}, top left).

To predict extreme events, we first compute $Y_0$ by convolving the field $u$ with a Gaussian with the length scale $L_G$ tuned to the particular spectrum.  After computing $Y_0$, we then compute the probability of an upcoming extreme event via our pre-computed conditional statistics (Figure~\ref{fig:condY0}, right).  We predict that an extreme event will occur if $P_{EE} > P^*$, where $P^*$ is a threshold probability that we choose.  Choosing a large $P^*$ will result in few false positives and many false negatives, while choosing a small $P^*$ will result in few false negatives and many false positives.  We choose $P^*=0.5$ as it gives a low rate of false negatives (meaning we predict almost all extreme events) with a reasonably low false positive rate.

\begin{figure}[H]
	\centering
	\includegraphics[width=\textwidth]{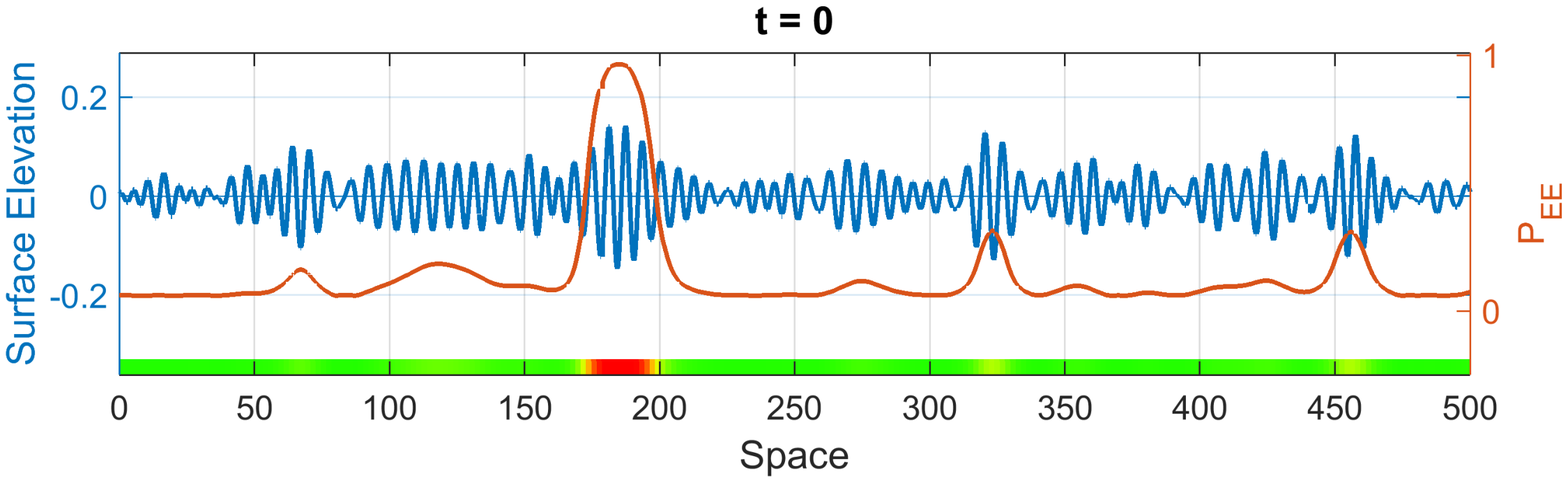} \\ [1em]
	\includegraphics[width=\textwidth]{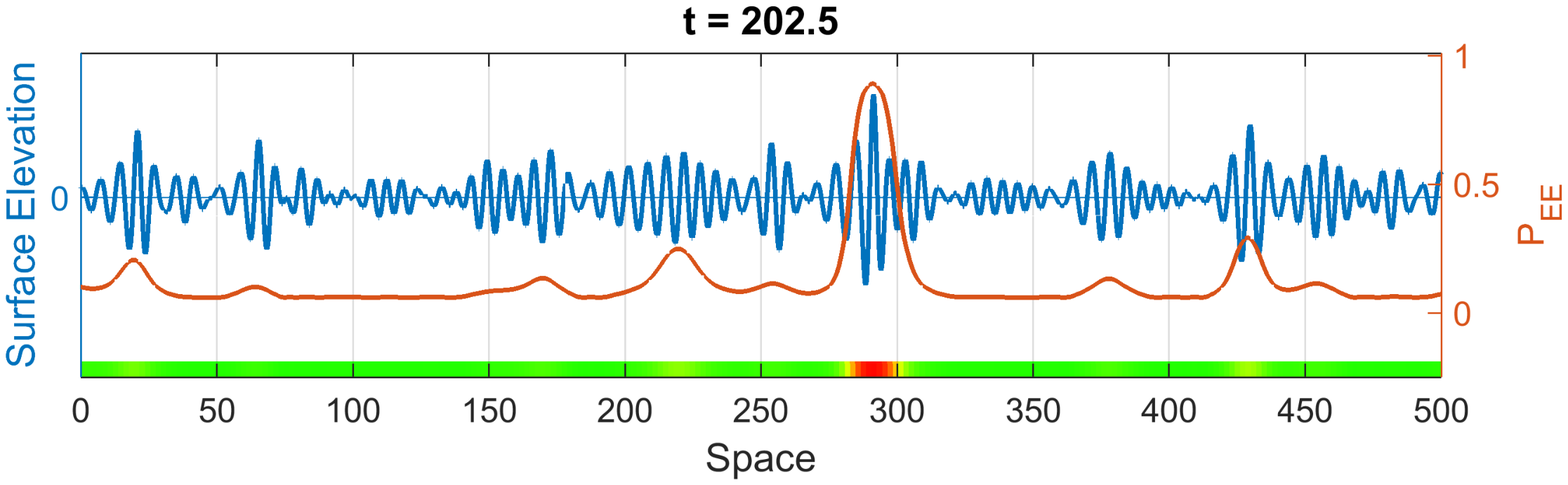} \\ [1em]
	\includegraphics[width=\textwidth]{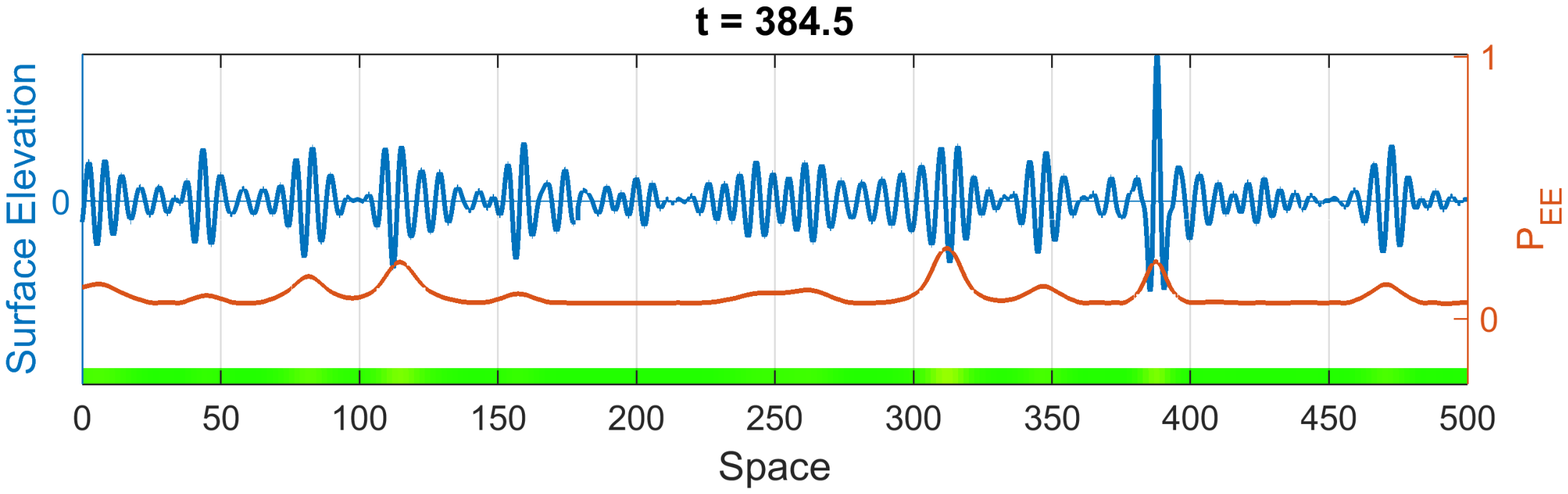}
	\caption{Surface elevation (blue) and probability of upcoming extreme event (red) for $t=0$ (top), $t=202.5$ (middle), and time at which maximum elevation attained, $t=384.5$ (bottom).}
	\label{fig:gaborExample}
\end{figure}

In Figure~\ref{fig:gaborExample}, we display the output of the Gabor predictive scheme applied to an example extreme event in a simulation of MNLS (this is the same example we investigated for the group detection-based predictive scheme in Figure~\ref{fig:multiscaleFS}.  The extreme event that occurs at $x\approx 390, t\approx 385$ is predicted from the initial conditions by our Gabor-based predictive scheme, with a high level of confidence ($P_{EE} = 0.965$).  This prediction occurs 200 time units (32 wave periods) before the amplitude exceeds $H_E$, and 385 time units (61 wave periods) before the maximally focused amplitude. The spatial location of the upcoming extreme event (in a moving coordinate frame) is predicted with very good accuracy as well.

To assess the reliability of this scheme we tested it on the 100 NLS and MNLS simulations used to test the group detection-based scheme in the previous section.  None of these simulations were used to generate the conditional statistics in Figure~\ref{fig:condY0}. We give data describing the performance of both predictive schemes in Table~\ref{tab:predStats}. The Gabor scheme predicted 316 of the 336 extreme events giving a low false negative rate of 5.9\%.  There were 108 instances where the Gabor scheme predicted an extreme event but one did not occur, giving a false positive rate of 25.5\%.  Thus, the Gabor scheme has a slightly higher false positive/negative rate compared with the group detection scheme but overall it performs nearly as well while requiring extremely little computational effort. Additionally, the average warning time for the Gabor scheme was larger than the group detection scheme (245 vs 153 time units, 39 vs 24 temporal wave periods).

\begin{table}
	\centering
	\begin{tabular}{c c c c c}
		Predictive Scheme & Extreme Events & False Neg. & False Pos. & Avg. Warning Time \\ \hline
		Gabor & 336 & 20 (5.9\%) & 108 (25.5\%) & 245 (39 periods)\\
		Group Detection & 336 & 0 & 91 (21.3\%) & 153 (24 periods)
	\end{tabular}
	\caption{Performance of Gabor and group detection prediction schemes on 100 simulations of MNLS.}
	\label{tab:predStats}
\end{table}

\section{Discussion}
\label{sec:discussion}

We have demonstrated how phase uncertainty (induced by the dispersive mixing of waves) and strong local nonlinearity interact with each other in nonlinear water waves resulting in the triggering of extreme waves. More specifically, we formulated a predictive scheme for rare events based on i) a nonlinear stability result that quantifies the focusing of localized wave groups and ii) the detection of wave groups that satisfy this localized stability criterion using a scale-selection algorithm applied much earlier, when the wave group still shows no obvious features that indicate a rare event. 

In a second stage we utilized the scale-selection algorithm to quantify, for a given wave spectrum, the probability for the formation of critical wave groups that can evolve into rare events. The combined analysis revealed a spatial length scale that has the highest probability to `trigger' a rare event. Based on this result we formulated an even simpler predictor that tracks the energy which can randomly accumulate over this length scale. We applied the two predictive schemes in the MNLS equation to forecast rare events in directional water waves. The results indicate very high accuracy of prediction, as well as robustness to the background wavenumbers, allowing for reliable predictions of, on average, 25-39 wave periods before the occurrence of an extreme wave.  

The presented approach introduces a new paradigm for understanding and predicting intermittent and localized events in dynamical systems characterized by important uncertainty and potentially strong nonlinear mechanisms. Future efforts include extension to two-dimensional water waves, study of the effects of bathymetry, as well as combination of the presented approach with standard filtering schemes formulated for linear and weakly non-linear systems in order to extend even further the prediction window.

\section*{Acknowledgments}
This research has been partially supported by the Naval Engineering Education Center (NEEC) grant 3002883706 and by the Office of Naval Research (ONR) grant ONR N00014-14-1-0520. The authors thank Dr. Craig Merrill (NEEC Technical Point of Contact) for numerous stimulating discussions.

\appendix
\section{Scale Selection Algorithm}
\label{app:scaleSelection}

Here we describe the wave group detection algorithm used in the prediction scheme discussed in Section~\ref{sec:prediction}\ref{sec:multiscale}. To find the dominant wave groups in a given irregular wave field, we look for Gaussian-like ``blobs" in $|u(x)|$.  To find these blobs, we use an existing algorithm based on the scale normalized derivatives of $|u|$ \cite{koenderink1984,witkin1984,lindeberg1998}. These scale-normalized derivatives, $s^{(m)}$, are normalized spatial derivatives of the convolution of $|u|$ with the heat kernel:

\begin{align*}
	s^{(m)}(x,L) = L^{m/2} \frac{\partial^m}{\partial x^m} \left( f \ast g \right)
\end{align*}
where $g(x,L)$ is the heat kernel:
\begin{align*}
	g(x,L) = \frac{1}{\sqrt{2\pi L}} e^{-\frac{x^2}{2L}}.
\end{align*}
in the above, $x$ is the spatial variable and $L$ is the length scale variable. Following \cite{lindeberg1998} we choose $m = 2$ for optimal blob detection.

We illustrate this approach by computing the scale-normalized derivatives of a single Gaussian: $u(x) = A e^{-x^2/2L_0^2}$.  It is straightforward to show that $s^{(2)}$ has a local minimum at $x=0, L = 2 L_0^2$.  In an arbitrary field $|u|$, we similarly find wave groups by computing $s^{(2)}$ and subsequently find the local minima.  If we find a local minima at $(x_C,L^*)$, we conclude that there is a wave group at $x=x_C$ having a length scale $\sqrt{L^*/2}$.  We determine the amplitude of the wave group, $A$, by computing the local maxima of $|u|$ near $x_C$.

\begin{figure}[H]
	\centering
	\includegraphics[width=0.45\textwidth]{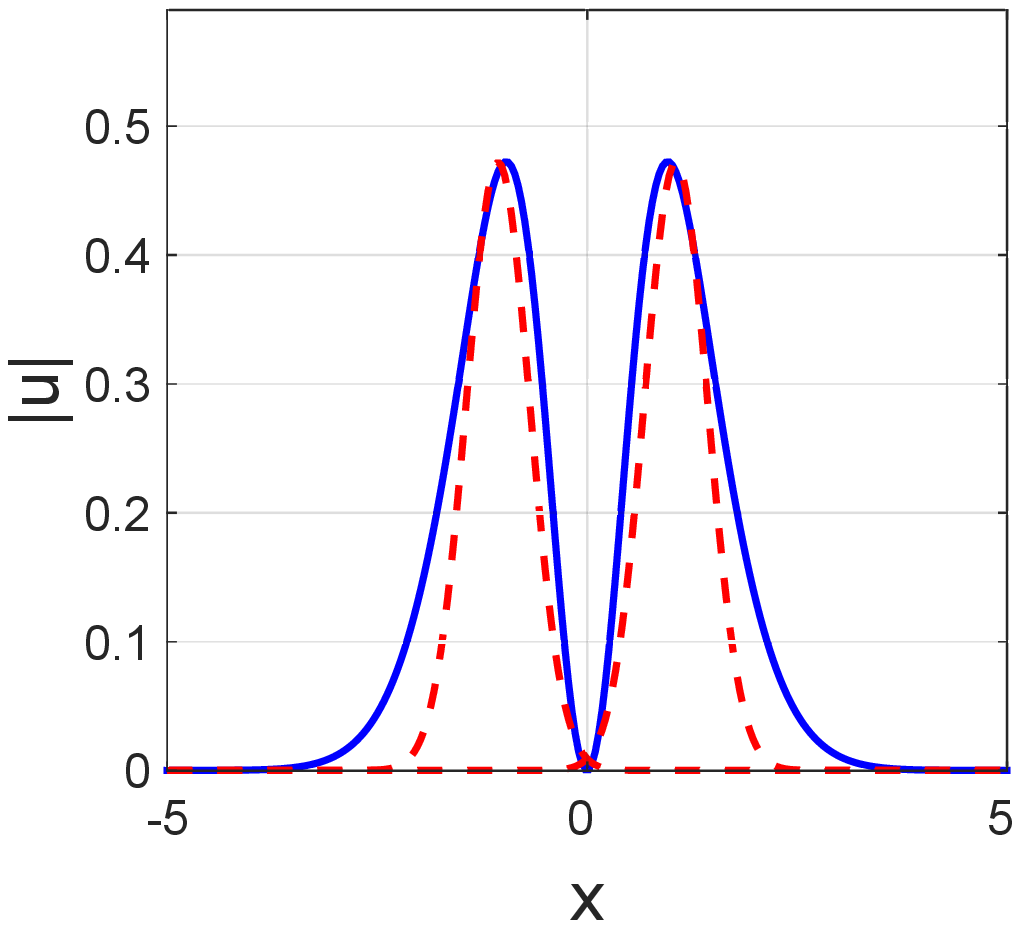} \includegraphics[width=0.45\textwidth]{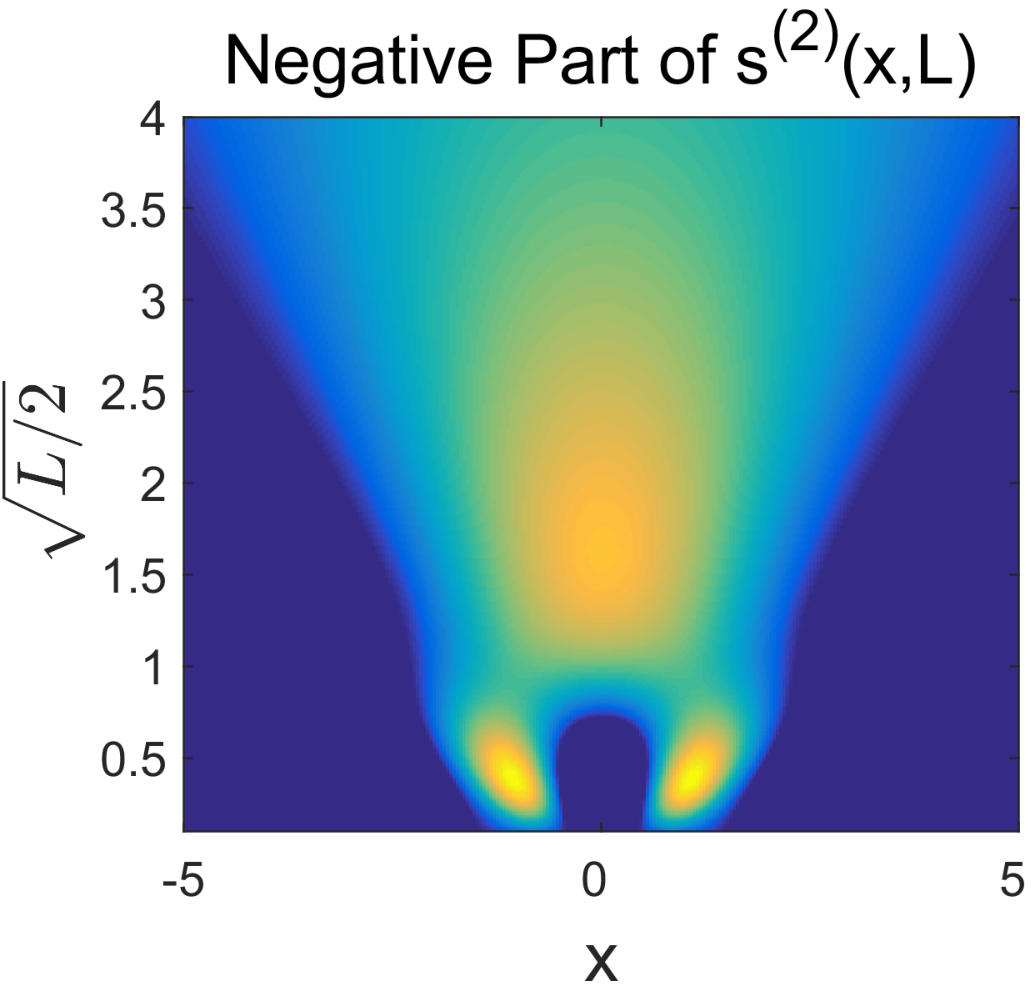} 
	\caption{Left: Plot of double-humped field $|u|$ and the groups identified via the scale selection algorithm. Right: Negative part of $s^{(2)}$.  The extremum at $x=0, \sqrt{L/2} = 2$ is eliminated via the procedure described in the text.}
	\label{fig:SSTwoHump}
\end{figure}

In some instances, scale space extrema do not correspond to actual wave groups. Consider $u(x) = e^{-x^2/2L_0^2} - e^{-x^2/2L_1^2}$, where $L_1 = 0.8L_0$, displayed in the left pane of Figure~\ref{fig:SSTwoHump}. There are two distinct peaks, which the scale selector detects.  However, there is an additional local minimum of $s^{(2)}$ near $x=0$ that does not correspond to a wave group.  To eliminate such false positives, for each local minimum of $s_2$, we compute the quantity $C$, which measures how close $u$ is to a Gaussian-like blob:

\begin{align*}
  C = 1 - \frac{||f(x)-A e^{-(x-x_C)^2/2(L^*)^2} ||_2}{||A e^{-(x-x_C)^2/2(L^*)^2} ||_2}
\end{align*}
if $|u|$ is an exact Gaussian, then $C$ is 1.  Thus, we take small values of $C$ as evidence that the local minimum of $s^{(2)}$ do not correspond to a wave group.  In the two-humped case displayed in Figure~\ref{fig:SSTwoHump}, the two local minima around $x=-1,1$ have $C\approx0.9$, while $C\approx0.5$ at $x=0$.  In practice, we ignore local minima of $s^{(2)}$ where $C<0.75$.  An example output of the scale selection algorithm with this criteria applied to an irregular wave field is displayed in the top pane of Figure~\ref{fig:groupStatistics}.  We see that the algorithm successfully identifies the dominant wave group in the field.

To compute the local minima of $s^{(2)}$, we first generate initial guesses for the minima by computing $s^{(2)}$ on a grid with $N_{x,SS}$ spatial points and $N_L$ points in the length scale dimension.  We then refine the local minima of the grid-evaluated $s^{(2)}$ by Newton's method.  For a particular length scale value $L$, computation of of $s^{(2)}(\cdot,L)$ requires two fast Fourier transforms.  Thus, the cost is $\mathcal{O}(N_L N_{x,SS} \log N_{x,SS})$.  We have found that a relatively small $N_L$ is adequate to generate reliable initial guesses for the minima (we use $N_L=20$).

For the Newton iteration, we compute the gradient and Hessian of $s^{(2)}$ analytically (these analytic expressions do contain integral terms, which we evaluate numerically).  For each wave group, we refine ($x_C,L$) to five digits of precision, which requires no more than 3 Newton iterations in almost all cases. Note that compared to the grid computation of $s^{(2)}$, the cost of the Newton iterations is low.  The reason for this is that at each iteration the gradient and Hessian of $s^{(2)}$ need only be calculated at a single point.  This means that the associated integration is only over a small subset of the full spatial domain.

Prediction via this scale selection algorithm is considerably cheaper than solving the envelope PDE.  In the example considered in Figure~\ref{sec:prediction}\ref{sec:multiscale},  we predict an extreme event 200 time units in advance using the scale selection-based algorithm.  Evolving the field this many time units with the PDE would require thousands  of time steps, each costing $\mathcal{O}(N_{x,PDE} \log N_{x,PDE})$ to compute the nonlinear terms, where $N_{x,PDE}$ is the number of spatial grid points in the numerical PDE solver.  By contrast, the scale selection algorithm only requires $N_L=20$ evaluations of $s^{(2)}(\cdot,L)$, with each evaluation of $s^{(2)}(\cdot,L)$ requiring $\mathcal{O}(N_{x,SS} \log N_{x,SS}$) operations. To accurately resolve the small scale dynamics and the nonlinear terms, $N_{x,PDE}$ must be considerably greater than $N_{x,SS}$, which demonstrates clearly the computational gain of the proposed approach (for the considered setting we found that $N_{x,SS}$ can be 16 times smaller than $N_{x,PDE}$ with no loss of reliability.
     
\bibliographystyle{plain}
\bibliography{waveRefs}

\end{document}